\documentclass[twocolumn]{aastex7}
\usepackage{hyperref}
\usepackage{printlen}
\graphicspath{{figures/}}

\begin{document}

\title{High-Resolution Dayside Spectroscopy of the Ultra-Hot Jupiter WASP-178b \\with GHOST/Gemini South}

\author[orcid=0009-0009-9314-6263,gname='Caitlin',sname='Bell']{Caitlin E.~Bell}
\affiliation{Department of Physics and Astronomy, East Texas A\&M University, Commerce, TX, USA}
\affiliation{International Gemini Observatory/NSF NOIRLab, Casilla 603, La Serena, Chile}
\email{artistcaitlin@gmail.com}  

\author[0000-0001-9796-2158,gname='Emily',sname='Deibert']{Emily K.~Deibert}
\affiliation{Department of Physics and Astronomy, University of Waterloo, 200 University Avenue West, Waterloo, Ontario N2L 3G1, Canada}
\affiliation{Waterloo Centre for Astrophysics, University of Waterloo, Waterloo, Ontario N2L 3G1, Canada}
\affiliation{International Gemini Observatory/NSF NOIRLab, Casilla 603, La Serena, Chile}
\email{emily.deibert@uwaterloo.ca}

\author[0000-0002-4451-1705]{Adam B. Langeveld}
\affiliation{School of Earth and Space Exploration, Arizona State University, Tempe, AZ 85287, USA}
\email{alangev1@jh.edu}

\author[0000-0001-6391-9266]{Ernst de Mooij}
\affiliation{Astrophysics Research Centre, Queen’s University Belfast, Belfast BT7 1NN, UK}
\email{e.demooij@qub.ac.uk}

\author[0000-0001-5349-6853]{Ray Jayawardhana}
\affiliation{Department of Astronomy, California Institute of Technology, Pasadena, CA 91125, USA}
\email{rayjay@caltech.edu}

\author[0000-0001-6362-0571]{Laura Flagg}
\affiliation{Department of Astronomy \& Astrophysics, 525 Davey Laboratory, The Pennsylvania State University, University Park, PA 16802, USA}
\affiliation{Center for Exoplanets and Habitable Worlds, 525 Davey Laboratory, The Pennsylvania State University, University Park, PA 16802, USA}
\email{laura.s.flagg@gmail.com}

\author[0000-0001-7836-1787]{Jake D. Turner}
\affiliation{Department of Astronomy and Carl Sagan Institute, Cornell University, Ithaca, NY 14850, USA}
\email{jaketurner@cornell.edu}

\begin{abstract}
    We present high-resolution dayside spectroscopy of the ultra-hot Jupiter WASP-178b obtained with the Gemini High-resolution Optical SpecTrograph (GHOST) at the Gemini South Observatory. The observations cover pre- and post-eclipse orbital phases, lasting approximately 9 hours in total, and represent the first published high-resolution optical dayside emission spectroscopy of WASP-178b's atmosphere, complementing previous near-infrared dayside spectroscopy. We verify the presence of a thermal inversion in the dayside atmosphere with a high-significance ({8.3}$\sigma$) detection of neutral iron emission via the Doppler cross-correlation technique, alongside hints of neutral Si and Ca emission. We also carry out model injection/recovery tests for other atmospheric species, which indicate that we do not expect to detect the majority of species searched for in this work, assuming our models are accurate representations of the planet's atmosphere. Finally, we place our results into context with previous work, showing how our tentative detection of neutral Si complements existing space-based observations and adds a new data point to previous detections of Si in ultra-hot Jupiter atmospheres. Our high-resolution observations provide important context about the dayside of WASP-178b, shedding light on 3D atmospheric processes and the extreme conditions present in ultra-hot Jupiter atmospheres.
\end{abstract}


\section{Introduction} \label{sec:intro}

    Tidally locked ultra-hot Jupiters (UHJs; \citealt{Parmentier18,Bell_2018}) are exoplanets with equilibrium temperatures $\gtrsim$ 2200 K. They have permanently irradiated daysides reaching temperatures upward of 3000 K, causing molecules to dissociate and atomic species to ionize \citep[e.g.,][]{Parmentier18,Bell_2018}, allowing for the detection of numerous gaseous metals \citep[Fe, Mg, Ti, Ca${}^+$, etc.; e.g.,][among others]{Pelletier_2023,Prinoth23,Langeveld25}. In contrast, temperatures on their nightsides are much cooler, and these same species may recombine and/or condense out of the atmosphere entirely, potentially leading to cold-trapping \citep[e.g.,][]{Pelletier_2023}. Dynamical processes can link the atmospheric conditions of these differing hemispheres, depending on their efficiency. Thus, UHJs allow us to explore atmospheric physics and chemistry under extreme conditions, and high-resolution spectroscopy---with its ability to resolve many individual spectral lines-- offers a particularly powerful probe of their atmospheres.
        
    The UHJ WASP-178b \citep{Hellier_2019} has an equilibrium temperature \(>\) 2400 K and an orbital period of 3.34 d \citep{Mart_nez_2020}. Its host star, WASP-178/HD 134004 (V = 9.95, spectral type A1 IV-V) has a T\textsubscript{eff} $\sim$ 9210 K \citep{Hellier_2019}, making it the second hottest star known to host a UHJ (just behind KELT-9 with T\textsubscript{eff} $\sim$ 10170 K, \citealt{Gaudi_2017}; and ahead of KELT-20/MASCARA-2 with T\textsubscript{eff} $\sim$ 8730 K, \citealt{Gaudi_2017,Talens_2018}). As such, WASP-178b is ideally suited to atmospheric characterization via high-resolution spectroscopy, and has been the subject of atmospheric investigations since its discovery in 2019 by the WASP-South survey, reported by \cite{Hellier_2019}. 

    The atmosphere of WASP-178b has previously been investigated in transit using both the Hubble Space Telescope (HST) and the James Webb Space Telescope (JWST). Based on HST data collected with WFC3/UVIS using G280, \cite{Lothringer22} reported strong near-ultraviolet (UV) absorption in the planet's atmosphere with a significance of $18\sigma$, attributing this absorption to silicate condensates and suggesting the presence of SiO and/or Mg and Fe${}^+$ in the terminator region of the atmosphere. As part of their analysis, they also investigated the presence of Fe, Fe${}^+$, TiO, and VO, all of which yielded non-detections. Fe and Fe${}^+$ did not provide significant contribution to the spectrum, while TiO and VO abundances were negligible and not a significant contributor to the near-UV absorption feature. 
    
    Subsequently, new observations were conducted with HST through WFC3/UVIS using G102 and G141, which were combined with JWST NIRSpec G395H observations to investigate carbon and oxygen abundances in WASP-178b's atmosphere \citep{Lothringer_2025}. The authors attributed the strong near-UV absorption to Mg and Fe${}^+$ with possible contributions from SiO, and weak IR features of H$_2$O and CO were identified. No evidence for Si or CO$_2$ in the IR was reported, while abundance inferences remained model dependent. \cite{Saha25} reanalyzed the JWST NIRSpec/G395H transmission spectra, reporting detections of CO (7.24$\sigma$) and CO${}_2$ (7.22$\sigma$), marginal evidence for C${}_2$H${}_2$ (1.34$\sigma$), and a super-solar carbon-to-oxygen ratio.
    
    At high spectral resolution, further investigations into WASP-178b's atmosphere were conducted with the Echelle SPectrograph for Rocky Exoplanets and Stable Spectroscopic Observations (ESPRESSO), where \cite{Damasceno2024} reported quantitative detections of refractory species over the course of two transits. These included single-line detections of Na, 
    H-$\alpha$,  
    H-$\beta$,  
    as well as cross-correlation detections of Fe,  
    and tentatively Mg 
    (as even with significance \(>\) 5$\sigma$, there were disparities in the $K_p$ values in the second night's detection). They further reported non-detections of Li (possibly due to low SNR) as well as H-$\gamma$. 

    In addition to being extensively characterized via transmission spectroscopy, WASP-178b's dayside atmosphere has been investigated with the Cryogenic high-resolution InfraRed Echelle Spectrograph (CRIRES$^+$) by \cite{cont2024}, who reported significant detections of $^{12}$CO (8.9$\sigma$) and H$_2$O (4.9$\sigma$), alongside non-detections of $^{13}$CO, OH, and Fe due to a lack of prominent lines within CRIRES$^+$ wavelength range.
    
    More recently, \cite{Fossati25} extended the characterization of WASP-178b by modeling its full vertical temperature–pressure (T-P) structure. They presented the first Non-Local Thermodynamic Equilibrium (NLTE) simulations of the planet’s atmosphere, generating synthetic transmission spectra that reproduce the observed near-UV absorption without requiring SiO. Their results show that Fe${}^+$ heating and Mg cooling drive a strong thermal inversion, rising from $\sim$2200 K at depth ($\sim$10 mbar) to over 8000 K in the upper atmosphere ($\sim10^{-8}$ bar). This approach follows similar NLTE analyses previously applied to KELT-9 b \citep[][]{Fossati20,Fossati21}, where it is important to acknowledge how the intense near-UV irradiation created by these planets' A-type stars can alter the ionization balance of metals \citep{Gaudi_2017}. However, they suggest that an NLTE model may be, surprisingly, an overestimate for observations in the optical, and focus their comparisons on transmission spectroscopy, as most of the retrievals done on WASP-178b's atmosphere have been done in transmission. We note that our analysis adopts an LTE framework, and focuses on the optical-to-near-IR (383 -- 1000 nm) emission spectrum.

    In this paper, we present high-resolution dayside spectroscopy of WASP-178b obtained with the Gemini High-resolution Optical SpecTrograph (GHOST; \citealt{GHOST,Kalari24}) at the Gemini South Observatory in Chile. Our aim is to characterize the dayside atmosphere of WASP-178b in the optical at pre- and post-eclipse orbital phases. We present the first neutral iron detection at high significance in the dayside atmosphere of WASP-178b, consistent with the Fe detection in the planet's transmission spectrum presented by \cite{Damasceno2024}. Our analysis further includes an investigation into the presence of other species in WASP-178b's atmosphere, as well as the results of model-injection/recovery tests. In Section \ref{sec:observations}, we describe the observations presented in this work. Our data reduction process is described in Section \ref{sec:reduction}, while the methods used to generate our atmospheric models are presented in Section \ref{sec:models}. We present our analysis and results in Sections \ref{sec:analysis} and \ref{sec:results} respectively, with a discussion following in Section \ref{sec:discussion}. Finally, we present our conclusions in Section \ref{sec:conclusions}.

\section{Observations} \label{sec:observations}

\begin{figure}
    \centering
    \includegraphics[width=\linewidth]{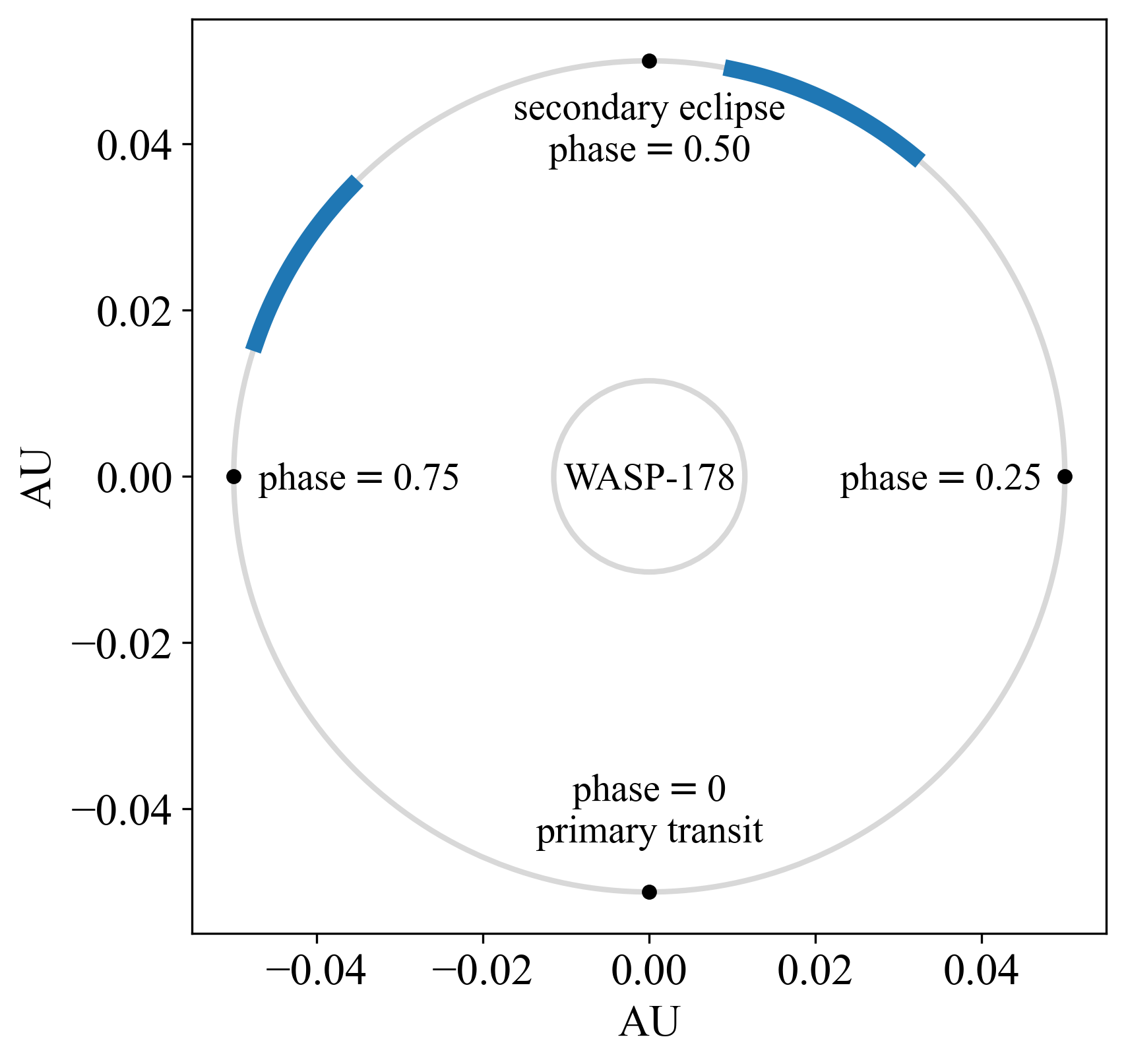}
   \caption{A visualization of the pre- and post-eclipse orbital phases (night 1 and night 2 respectively) covered by these observations. The star, planet, and orbit are drawn to scale.}
    \label{fig:orbit diagram}
\end{figure}

    We obtained two phase-curve observations of WASP-178b over the course of two nights with a cumulative duration of approximately 9 h using GHOST at the Gemini South Observatory in Chile. The observations are summarized in Table~\ref{tab:observations}. In total, 272 spectra were captured, targeting the planet's dayside atmosphere before and after secondary eclipse at orbital phases of $\sim$0.39 to $\sim$0.47 (136 exposures) and $\sim$0.63 to $\sim$0.70 (136 exposures) respectively (see Figure~\ref{fig:orbit diagram}). We calculated this orbital phase coverage using the planetary system parameters presented in Table~\ref{tab:parameters}.

    \begin{deluxetable*}{lccccc}
\tablewidth{0pt}
\tablecaption{Planetary system parameters used in this work.  \label{tab:parameters}}
\tablehead{
\colhead{Parameter} & \colhead{Symbol [unit]} & \colhead{Value} &\colhead{Reference} }
\startdata
Transit Midpoint & $T_0$ [BJD] & $2458321.86724_{-0.00039}^{+0.00038}$ & \cite{Mart_nez_2020} \\
Period & $p$ [days] & $3.3448412\pm{0.0000033}$ & \cite{Mart_nez_2020} \\
Planetary Radius & $R_p$ [$R_J$] & $1.940_{-0.058}^{+0.060}$ & \cite{Mart_nez_2020} \\
Planetary Mass & $M_p$ [$M_J$] & $1.41_{-0.51}^{+0.43}$ & \cite{Mart_nez_2020} \\
Planetary Equilibrium  Temperature & $T_{eq}$ [$K$]& $2402 \pm 60$ & \cite{Hellier_2019} \\
Host Star Spectral Type & -- & A1 IV-V & \cite{Hellier_2019} \\
Stellar Radius & $R_*$ [$R_{Sun}$] & $1.67\pm{0.07}$ & \cite{Hellier_2019} \\
Stellar Effective Temperature & $T_*$ [$K$] & $9360\pm{150}$ & \cite{Hellier_2019} \\
Systemic Velocity & RV${}_{sys}$ [km/s] & $-23.908\pm{0.007}$ & \cite{Hellier_2019} \\
Planetary Keplerian Velocity & $K_p$ [km/s] & 176.5\tablenotemark{a} & \cite{cont2024} \\
T-P Profile Temperature Point 1 & $T_1$ [K] & 3661 & \cite{cont2024} \\
T-P Profile Pressure Point 1 & $\log p_1$ & $-4.62^{+1.56}_{-1.88}$ & \cite{cont2024} \\
T-P Profile Temperature Point 2 & $T_2$ [K] & 2756 & \cite{cont2024} \\
T-P Profile Pressure Point 2 & $\log p_2$ & $-0.33^{+1.64}_{-1.53}$ & \cite{cont2024} \\
\enddata
\tablenotetext{a}{Calculated by \cite{cont2024} via $K_p = (2\pi ^ M_*/P_\mathrm{orb})^{1/3}$}
\end{deluxetable*}

    As in \cite{Deibert24}, we set the instrument binning to 1 x 4 (spectral x spatial), and used the high-resolution, single-object observing mode with IFU1 (which has a total area on the sky of 0.92 arcsec${}^2$) placed on the target and the dedicated sky IFU, IFU2 (which has a total area on the sky of 0.34 arcsec${}^2$), placed on the sky. The exposure time was set to 120s per spectrum in both the red and blue cameras, and as in \cite{Deibert24}, we configured the instrument such that subsequent red and blue exposures would begin at the same time. Unlike in \cite{Deibert24}, we used the ``slow'' read mode for the blue camera and the ``medium'' read mode for the red camera, as this is now the suggested read mode combination for most GHOST observations according to the GHOST instrument web pages\footnote{\url{https://www.gemini.edu/instrumentation/ghost/observation-preparation}}, due to the fact that these settings yield read times in both cameras of roughly the same length. In particular, this setup results in a readout time of 14.4s in the blue camera and 13.9s in the red camera, with the overall cadence of the observations thus being $\sim$134s.

    The first night of observations contained several short ($\sim$10 minute-long) gaps in the orbital phase coverage due to software issues with the instrument. However, the total observing time of $\sim$4.5 h made up of 136 exposures was consistent across both nights. These gaps in the orbital phase coverage did not impact our final analysis; they simply changed the exact orbital phases covered by each spectrum. We note that the data presented in \cite{Deibert24} contained similar gaps in the orbital phase coverage. Future observations should not suffer from similar issues, as GHOST has now been reliably integrated into the Gemini South queue.

    \begin{deluxetable*}{lccccc}
\tablewidth{0pt}
\tablecaption{Summary of WASP-178 b GHOST/Gemini South Observations. \label{tab:observations}}
\tablehead{
\colhead{Date (UT)} & \colhead{Num. Exposures} & \colhead{Exposure Time (s)} & \colhead{Orbital Phase Coverage} & \colhead{Airmass Variation\tablenotemark{a}} & \colhead{SNR Variation\tablenotemark{b}} }
\startdata
2024 Mar 24 & 136 & 120 & 0.39 -- 0.47 & 1.79 -- 1.02 -- 1.08 & 82.02 -- 99.26 (red) \\
 & & & & & 76.94 -- 98.04 (blue) \\
2024 May 04 & 136 & 120 & 0.63 -- 0.70 & 1.41 -- 1.02 -- 1.13 & 53.05 -- 78.21 (red) \\
 & & & & & 47.64 -- 72.79 (blue) \\
\enddata
\tablenotetext{a}{Beginning -- Minimum -- End}
\tablenotetext{b}{Refers to the minimum and maximum values of the average SNR per pixel across all orders for the blue and red cameras. Note that this is not including orders which were excluded from the analysis due to being outside the usable GHOST wavelength range of 383 -- 1000 nm \citep{Kalari24}.}
\end{deluxetable*}

\begin{figure*}
    \centering
    \includegraphics[]{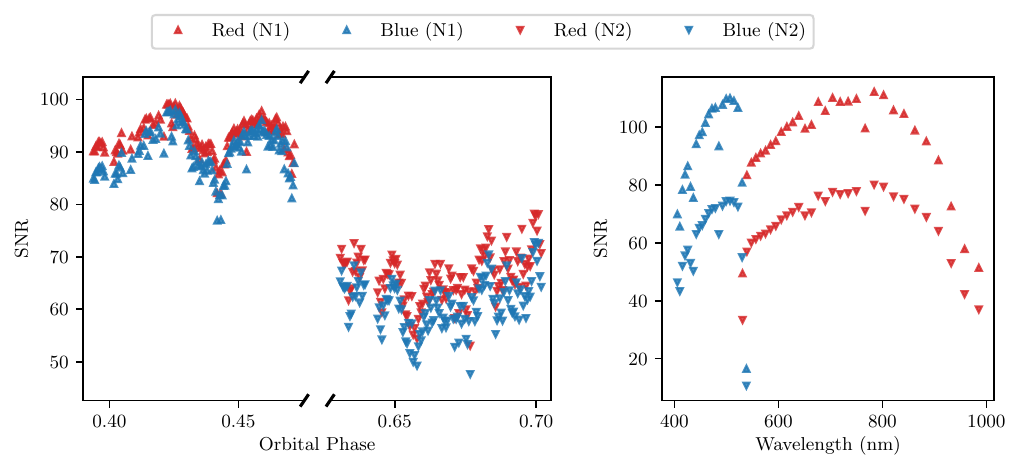}
    \caption{Left: The average SNR per exposure for both nights of our observations in both the blue and red cameras. Night 1, corresponding to pre-eclipse orbital phases, is plotted before the break in the x-axis; while night 2, corresponding to post-eclipse phases, is plotted after the break in the x-axis. Note that the break in the x-axis does not correspond exactly to secondary eclipse. Upward-pointing triangles indicate the night 1 data, while downward-pointing triangles indicate night 2 data.
    Right: The average SNR per order for both nights, plotted against the median wavelength of each order. As in the subplot on the left, the red and blue cameras are color-coded in red and blue, while night 1 is represented by upward-pointing triangles and night 2 by downward-pointing triangles.
    As is clear from the figure, the data quality was on average higher during the first night of observations (likely due to the better observing conditions as described in Section \ref{sec:observations}).}
    \label{fig:snr}
\end{figure*}

    The weather conditions remained generally stable throughout the course of the two observing periods. On night 1 (pre-eclipse), the sky remained mostly clear, and the seeing was around 0.7 to 0.8 arcseconds (i.e., within the IQ70 bin as defined at Gemini South). On night 2 (post-eclipse), some thin cirrus cloud cover was present throughout the observations, and the seeing was typically around 1 arcsecond (i.e., within the IQ85 bin). Further details on the signal-to-noise ratio (SNR) and airmass variation throughout the observations are presented in Table \ref{tab:observations}. The average SNR per exposure in both the blue and red, as well as the average SNR per order in both the blue and red, is also presented visually in Figure \ref{fig:snr}. As is clear from the figure, the SNR was higher on average on night 1; this is likely due to the better observing conditions as described earlier. Note as well that this does not include orders outside of the ``usable'' GHOST wavelength range ($\sim$383 -- 1000 nm; \citealt{Kalari24}). These orders (2 out of 33 in the red, and 13 out of 35 in the blue) were not used in our analysis.

    Raw and reduced data products for the observations presented in this work are available in the Gemini Observatory Archive (GOA) under the program ID GS-2024A-Q-138 (PI: Deibert). 
    
\section{Data Reduction} \label{sec:reduction}
\subsection{Initial Reduction} \label{subsec:initial_reduction}

We reduced the data using version 3.2.0 of the \texttt{DRAGONS} data reduction software \citep[][]{dragons19,dragons22}. This initial reduction includes a bias subtraction, flat-fielding, cosmic ray rejection, bad pixel masking, optimal extraction, and wavelength calibration. Optionally, the software can also be used to perform a sky subtraction and/or a barycentric correction. Via a visual inspection of the data, whereby we searched for known strong telluric emission lines (due e.g., to atomic oxygen at $\sim$557.7 nm and 630 nm, sodium at $\sim$589 nm and 589.6 nm, and hydroxyl bands in the redder end of the data), we determined that the sky subtraction was not necessary for our analysis. We therefore reduced the data with the sky subtraction and barycentric correction turned OFF, as we need the data in the telluric frame for our telluric removal process. We note that the default reduced data product provided by the Gemini South Observatory is produced with both the sky subtraction and the barycentric correction turned ON. We used the barycentric Earth radial velocity (BERV) values from these data products to later shift the data to the barycentric frame. The products available in the GOA include both of these corrections; however, the reduced products used in this work are available from the authors upon request (or can be produced by rerunning the \texttt{DRAGONS} software with the sky subtraction and barycentric correction turned OFF). 

After this initial correction, we largely followed the methods presented in \cite{Deibert24}. We first ran an additional outlier-masking routine to account for any outliers (e.g., cosmic rays) which were not flagged by the \texttt{DRAGONS} software. Any values which varied by more than 5 median absolute deviations were masked. We then median-normalized the spectra, in order to account for varying flux levels (due, for e.g., to varying observing conditions), and then subtracted a median frame, which provided an initial correction for stellar and telluric absorption.

\subsection{Telluric and Stellar Absorption Correction} \label{subsec:sysrem}

To correct for remaining stellar and telluric absorption features in the data, we made use of the \textsc{SysRem} algorithm \citep[][]{Tamuz2005}. \textsc{SysRem} is a Principal Component Analysis (PCA)-like algorithm which can be used to remove essentially time-stationary features (such as stellar and telluric absorption) in astronomical spectra. After converting the flux values to magnitudes, we ran the \textsc{SysRem} algorithm separately on the blue and red arms of the 2D data product from \texttt{DRAGONS} in the telluric rest frame, prior to any barycentric corrections. While the stellar features do have a slight Doppler shift, it is negligible for the purposes of the \textsc{SysRem} algorithm. We utilized the airmass to provide an initial guess for the first systematic to be removed, then ran the \textsc{SysRem} algorithm between one and 20 iterations order-by-order.

To determine the optimal number of iterations of the algorithm to use on our spectra, we made use of the $\Delta$CCF (cross-correlation function) method presented in \cite{Cheverall_2023}. Briefly, this method involves comparing a model-injected version of the data with the data itself, and using this to determine the number of \textsc{SysRem} iterations to apply to the data in an unbiased way. This method will result in a maximized SNR in the quantity $\Delta$CCF = $\mathrm{CCF}_{\text{inj}}$ - $\mathrm{CCF}_{\text{obs}}$, where $\mathrm{CCF}_{\text{inj}}$ and $\mathrm{CCF}_{\text{obs}}$ are the cross-correlation functions (CCFs) calculated via Doppler cross-correlation as described in Section \ref{subsec:doppler} for model-injected version of the data and the observed data, respectively. The calculation of the SNR values uses the $\Delta$CCF map directly, while the noise comes from $\mathrm{CCF}_{\text{obs}}$.

This method is robust against biases that may arise from optimizing \textsc{SysRem} on both $\mathrm{CCF}_{\text{inj}}$ and $\mathrm{CCF}_{\text{obs}}$ individually, as discussed by \cite{Cheverall_2023}. This method is also largely insensitive to different models and injection velocities used in the cross-correlation \citep{Cheverall_2023}. \cite{Cheverall_2023} suggest using a wide range of velocities in CCFs to limit biases in calculating the SNR.

While the $\Delta$CCF method can be applied on an order-by-order basis, optimizing each order for a different number of \textsc{SysRem} iterations, we instead carried out a global $\Delta$CCF optimization for the red and blue detectors and each night separately. To do this, we injected an Fe model (see Section \ref{sec:models}) into the data at the expected planetary velocity (using the value derived by \citealt{cont2024}; see Table \ref{tab:parameters}), and repeated our reduction and analysis process on this model-injected data for \textsc{SysRem} iterations ranging between one to 20. We chose to use Fe, as it is the only species detected at high significance, though we reiterate that \cite{Cheverall_2023} show that the results are consistent across models.

When measuring the planet's expected location, we found that the $\Delta$CCF detection significance was somewhat variable over the course of the 20 iterations, with several local peaks. We also note that there were several spurious $\sim3\sigma$ peaks at other locations in the data, including a persistent misplaced correlation peak in the results of both nights' $\Delta$CCF maps at $K_p \sim 70$ km/s and RV $\sim$ ${25}$ km/s respectively. Such spurious peaks at the $\sim3\sigma$ level are common in CCF maps, as demonstrated by e.g., \cite{Esteves17}. Throughout this work, we use $5\sigma$ as our threshold for confidently detecting a signal, in order to avoid treating spurious peaks as potential detections. 

\begin{figure*}
    \centering
    \includegraphics[width=\textwidth]{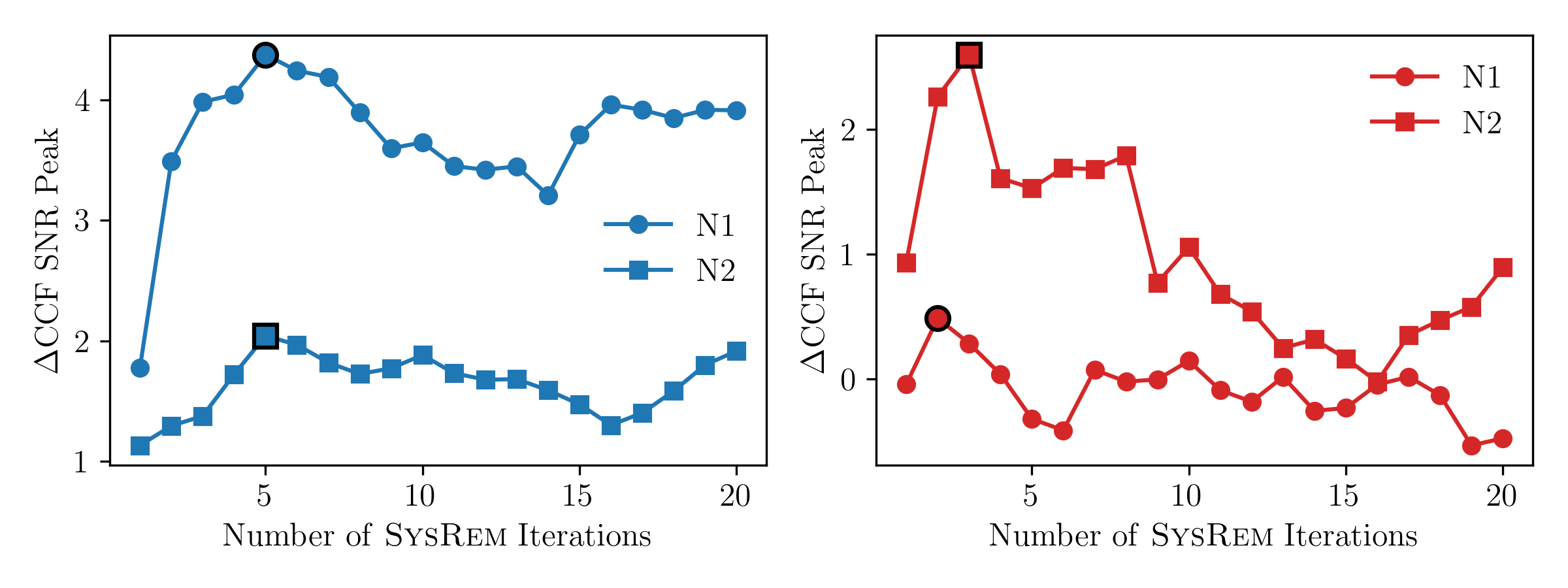}
    \caption{The results of applying the $\Delta$CCF method to the blue (left subplot) and red (right subplot) arms of the data for the two nights of our observations, as described in Section \ref{sec:reduction}. {The iteration used for each respective arm and night is indicated with a black outline.}}
    \label{fig:delta_ccf}
\end{figure*}

We chose to apply 2 and 3 iterations of the algorithm to nights 1 and 2 of the red camera data, respectively, corresponding to the peaks in detection significance determined via the $\Delta$CCF method (see Figure \ref{fig:delta_ccf}). {In the case of the blue camera data, we chose to apply 5 iterations of the algorithm to both nights corresponding to the peak of the detection significance}. While the red arm of the data contains significantly more telluric lines than the blue, the lower data quality of the blue arm could perhaps explain why additional iterations were needed in the blue. The root-mean-square (RMS) of the data continued to decrease with additional iterations of \textsc{SysRem} for both cameras on both nights, though likely at the expense of the planetary signal.

 We note that, as seen in Figure \ref{fig:delta_ccf}, the detection significance remained roughly consistent within 1$\sigma$ when any number of iterations of the \textsc{SysRem} algorithm between 1 to 10 was applied to the data. The exact number of iterations within this range therefore should not have a significant impact on our final results.

The results of applying the \textsc{SysRem} algorithm to our data are displayed in Appendix \ref{app:sysrem}. 
Note that there are still some residual telluric features and variations in the noise levels of the \textsc{SysRem}-corrected orders. While applying different numbers of iterations of the algorithm on a per-order basis may help further suppress these tellurics, we have opted to avoid additional fine-tuning, which may bias the results towards a positive detection.

In Section \ref{subsec:doppler}, we weight each pixel by its standard deviation to ensure that especially contaminated regions of the data will not contribute excess noise. This follows, for example, \cite{Snellen_2010}, and the standard deviation per-pixel can be seen in the final row of the figures in Appendix \ref{app:sysrem}.

\section{Atmospheric Models} \label{sec:models}

We modeled the atmosphere of WASP-178b using version 2.7.7 of \texttt{petitRADTRANS} \citep{petitradtrans}. Each model consisted of a single species embedded in an inert atmosphere, assuming solar abundances and metallicities. The system parameters used to generate these models are presented in Table \ref{tab:parameters}; the T-P profile was generated using the best-fit parameters from the retrieval analysis of a two-point T-P profile presented in \cite{cont2024}. In particular, we made use of their high-resolution-only retrieval parameters.

\begin{figure*}[ht!]
    \centering
    \includegraphics[width=\linewidth]{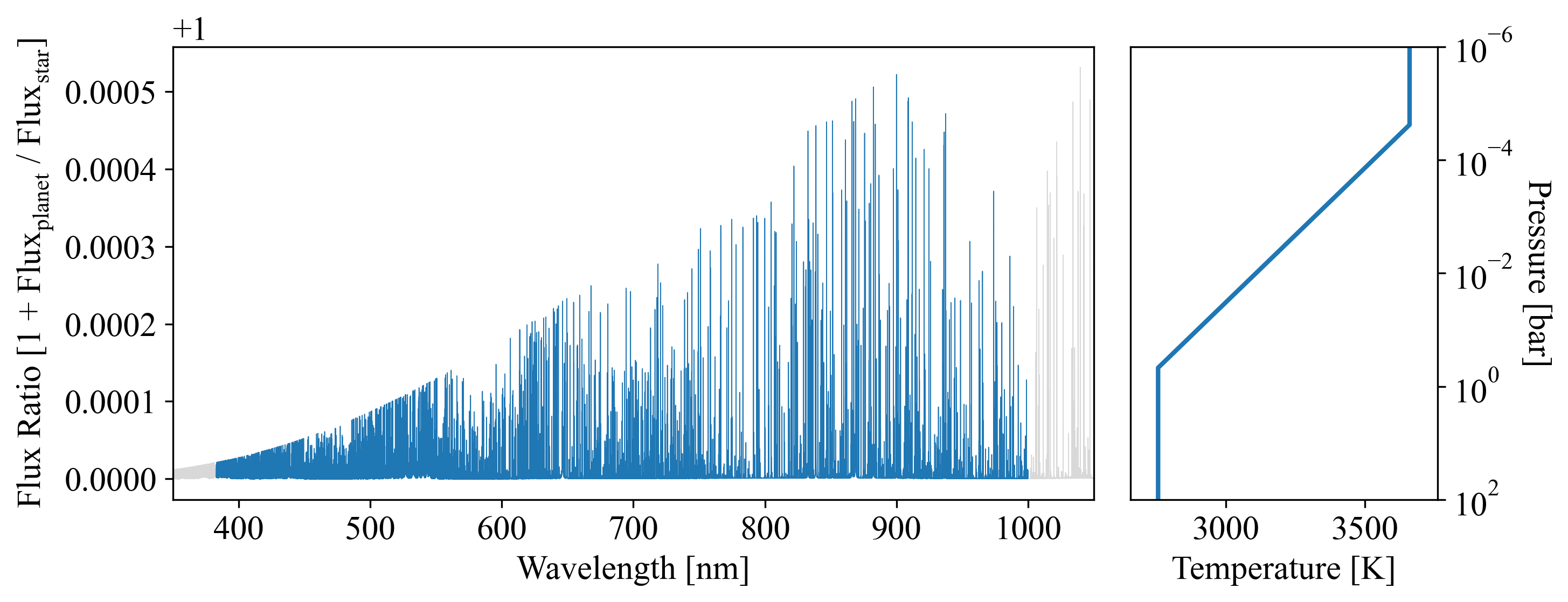}
    \caption{Left: The Fe model used in our analysis, generated with \texttt{petitRADTRANS} \citep{petitradtrans}, as described in Section \ref{sec:models}. The full GHOST wavelength is shown; however, we note that we used only a subset of this wavelength range (corresponding to the ``usable'' range of 383 -- 1000 nm; \citealt{Kalari24}) in our analysis, shown in blue. Right: The two-point \textit{T-P} profile used to generate the model, using the best-fit parameters from the high-resolution-only retrieval of \cite{cont2024} and the planetary parameters listed in Table~\ref{tab:parameters}.}
    \label{fig:Fe&TP}
\end{figure*}

We calculated abundances for each species using the \texttt{FastChem} chemical equilibrium model \citep{Stock18}. We used the volume mixing ratios (VMRs) as a function of pressure output by \texttt{FastChem} when generating our models with \texttt{petitRADTRANS}. Finally, we scaled the models by a blackbody of $T_* = 9360$ K, matching that of the host star \citep[][]{Hellier_2019}. Using these methods, we created individual models for Fe, Fe${}^+$, Ti, TiO, V, V${}^+$, VO, Al, Ca, Ca${}^+$, Cr, K, Mg, Na, Si, and FeH. The sources for each line list can be found in the Appendix of \cite{Deibert24} and references therein. An example model created for Fe, alongside the two-point T-P profile used to generate the model, is shown in Figure \ref{fig:Fe&TP}.

Before correlating the models with our data (see Section \ref{subsec:doppler}), we convolved them with an average instrumental profile matching the GHOST resolution, following the methods of \cite{Herman22}. Briefly, this involved using the \texttt{astropy.convolution.convolve} function, with a 1D Gaussian Kernel of standard deviation $\mathrm{FWHM}_\mathrm{GHOST} / 2.35 $, where the FWHM of GHOST's instrumental broadening in velocity space is assumed to be $\sim$3.9 km/s (assuming an average resolution of $\sim$76,000).

\section{Analysis} \label{sec:analysis}

We characterized the atmosphere of WASP-178b via a Doppler cross-correlation analysis \citep[e.g.,][]{Snellen_2010} with the high-resolution one-dimensional atmospheric models generated for WASP-178b's system parameters as described in the preceding section. Following the Doppler cross-correlation analysis, we also carried out model injection/recovery tests to better understand the limitations of our analysis. These methods are described in additional detail below.

\subsection{Doppler Cross-Correlation} \label{subsec:doppler}

After creating the models, we analyzed the data via the Doppler cross-correlation technique \citep[e.g.,][]{Snellen_2010}. To do this, we first shifted the data to the stellar rest frame using the BERV values calculated by the \texttt{DRAGONS} software, and the systemic velocity presented in Table \ref{tab:parameters}. We then Doppler-shifted the models to radial velocities (RVs) ranging from -300 km/s to +300 km/s, and cross-correlated the Doppler-shifted models with our data. This allowed us to create CCF maps for each model. {As described in Section \ref{subsec:sysrem}, the data have been weighted by their standard deviation (see the final row of the figures in Appendix \ref{app:sysrem}), ensuring that particularly contaminated regions of the data that may still contain residual noise following the \textsc{SysRem} correction do not contribute excess noise to the CCF maps.} 

To further boost the strength of any signals present in the data, we then phase-folded the CCFs to Keplerian velocities ($K_p$) ranging from 1 km/s to 300 km/s with step sizes of 1 km/s. In the final resultant $K_p-$RV map, a detection in the planet's atmosphere should be visible as a correlation peak close to the expected planetary Keplerian velocity and a radial velocity of zero (as we have shifted the data to the stellar rest frame). Any deviations from the expected location may indicate, for example, dynamics in the atmosphere.

We determined the significance of the $K_p-$RV maps by subtracting off the standard deviation of a 3$\sigma$-clipped version of the map (as in \citealt{Smith24}). We take 5$\sigma$ as our threshold for confidently detecting an atmospheric species.

\subsection{Model Injection/Recovery Tests} \label{subsec:injection}

In addition to the Doppler cross-correlation analysis described above, we also carried out model injection and recovery tests for each species. This involved injecting the model into the data at the negative of its expected Keplerian velocity, and then repeating the Doppler cross-correlation analysis. We injected the data at a negative velocity rather than the true expected Keplerian velocity in order to avoid artificially boosting any weak signals that are present in the data. This test allowed us to determine whether we should expect to detect a given model in our data (bearing in mind the fact that it will be easier to detect an artificially injected model than a true signal), or whether the data are not sufficient to detect a given model (due to the strength of the model's spectral lines and the noise level of the data).

\section{Results} \label{sec:results}

Following the methodology described in Section \ref{sec:analysis}, we cross-correlated our observations with model atmospheres generated for Fe, Fe$^+$, Cr, V, Ti, Ca, Ca$^+$, K, Si, Mg, Na, FeH, VO, and TiO. For most species, we carried this cross-correlation out for both the blue and red arms of the data. The exceptions are FeH, Ca+, and Si, for which there were no or few strong spectral lines present in the blue arm of the data. We note that all features in the CCF maps for night 2 appear more extended in velocity space than those in night 1, possibly due to the orbital phase coverage of night 2 (which is further from secondary eclipse than that of night 1), and/or the worse observation conditions of night 2 as discussed in Section \ref{sec:observations}. As a final step, we combine the results from night 1 and night 2, blue and red, {via a weighted average against each night/camera's respective average SNR (where the average SNR value is an average across all spectra for a particular night/camera)} (see Figure \ref{fig:snr}) to obtain a final detection (or tentative detection) for each species with constrained RV and $K_p$ values. Again, the exceptions were FeH, Ca$^+$, and Si, for which there were insufficient strong lines present in the blue.

\subsection{Neutral Iron Detection} 
\label{subsec:iron}

\begin{figure*}
    \centering
    \includegraphics[width=\linewidth]{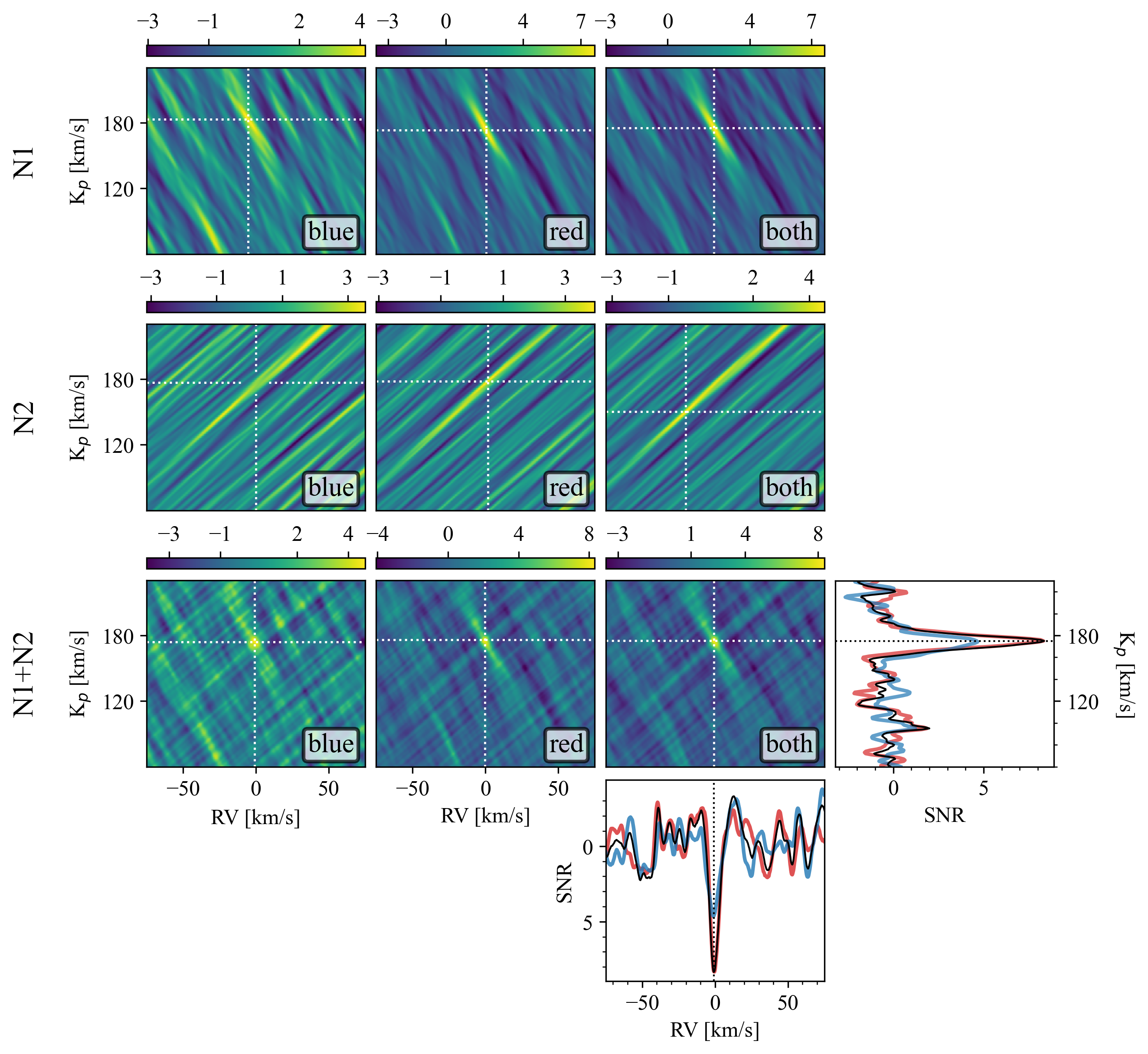}
    \caption{Row 1: The results of cross-correlating our night 1 data with the Fe model. {The location of the peak $K_p$ and RV values are indicated by converging dotted white lines in each plot with a significant detection. For plots without a significant detection, the expected planetary signal location is indicated by non-converging dotted white lines (i.e., RV $= 0$ km/s, $K_p \sim 176.5$ km/s).} The colorbars indicate the significances of the maps, determined via dividing out the standard deviation away from the peak signal, calculated via a 3$\sigma$-clipped map (see Section \ref{subsec:doppler}). Row 2: The same as row one, but for night 2. Row 3: The same as row 1, but for night 1 and 2 combined, with each night weighted by its respective average SNR. Surrounding plots labeled ``both'' are slices of the 2D $K_p-$RV maps at the peak $K_p$ and peak RV for the blue (blue line), red (red line), and combined (black line) results. The dashed black lines represent the combined peak RV and $K_p$.}
        \label{fig:Fe}
\end{figure*}

The results of cross-correlating our data with the atmospheric model generated for Fe (i.e., Figure. \ref{fig:Fe&TP}) are presented in Figure \ref{fig:Fe}. In the red, we detect Fe at significances of 7.7$\sigma$ and 4.0$\sigma$ for nights 1 and 2 respectively. In the blue arm, we detect Fe at a significance of 4.2$\sigma$ on night 1, and do not confidently detect it on night 2. We do note the presence of a 3.5$\sigma$ signal in the CCF map, however it is significantly offset from the expected planetary location. The combination of the blue and red data yields a detection significance of 7.6$\sigma$ and {4.5}$\sigma$ for each night respectively.

The final combination of the night 1 blue and red data with the night 2 blue and red data yields a final detection significance of {8.3}$\sigma$ detected at a Keplerian orbital velocity of {$K_p = 175.0^{+2.3}_{-2.6}$} km/s and a radial velocity of RV$ = -0.9^{+1.8}_{-1.3}$ km/s, where we have taken as error the 1$\sigma$ extent of the correlation peak. In Section \ref{sec:discussion}, we compare these results with previous detections in the atmosphere of WASP-178b.

\subsection{Tentative Detections \& Non-Detections}
\label{subsec:nondetections}

\begin{figure*}
     \centering
     \includegraphics[width=\linewidth]{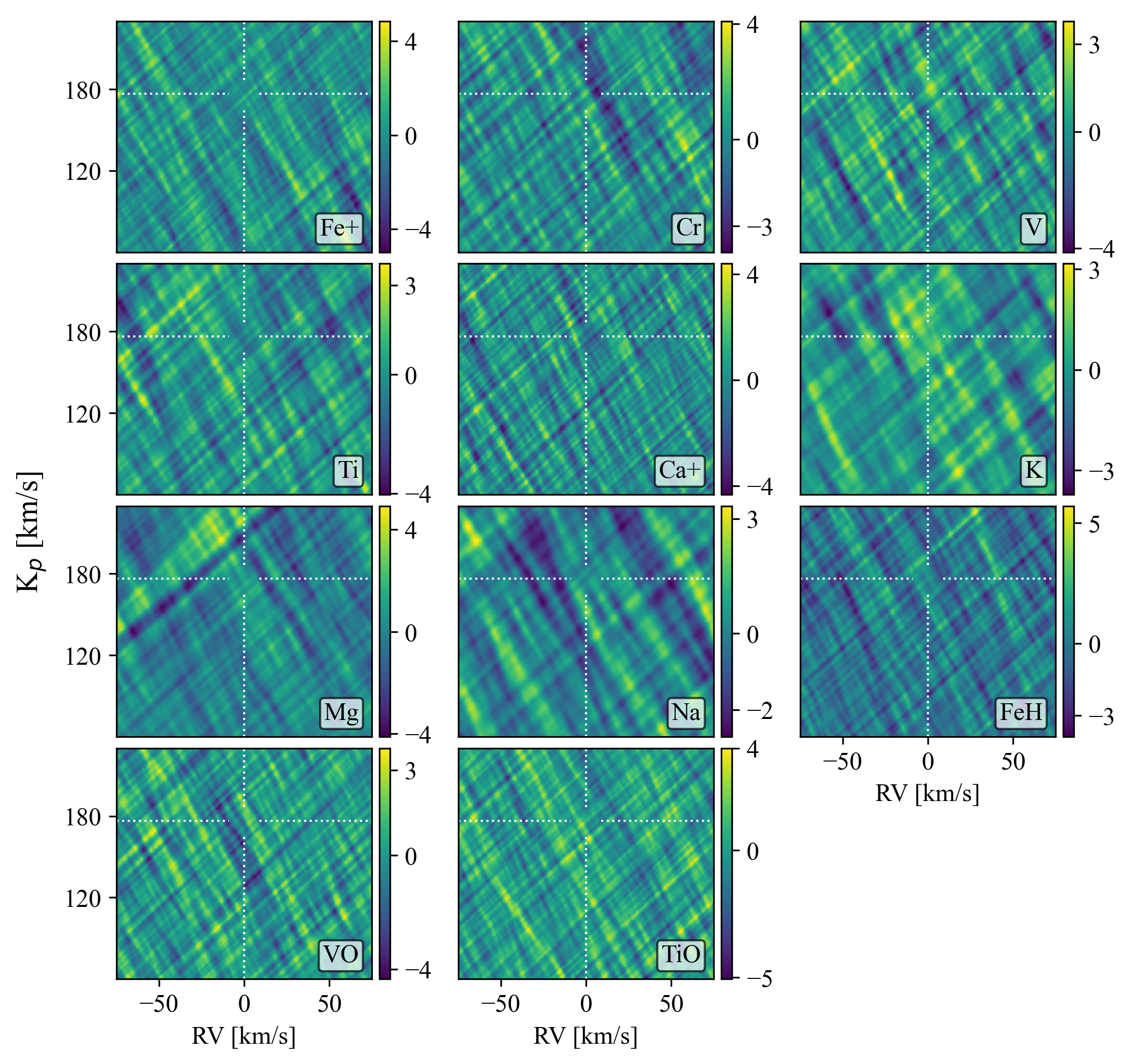}
     \caption{The combined night 1, night 2, blue (where applicable; see Section \ref{sec:results}), and red arm 2D $K_p-$RV maps for the species not detected in this work. The location of the expected planet location is indicated with {non-converging} dashed white lines (i.e., RV$=0$ km/s, $K_p\sim176.5$ km/s). The colorbars indicate the same as described in the caption of Figure \ref{fig:Fe}, and the species are indicated in the bottom-right corners.}
     \label{fig:nondetections}
 \end{figure*}

The remaining species in this work did not yield significant detections (i.e. $> 5\sigma$) in the blue, red, or combined results, for both nights 1 and 2 of our observations. We present the final combination of the night 1 blue and red data with the night 2 blue and red data in Figure \ref{fig:nondetections} for these undetected species. 

\begin{figure}
    \centering
    \includegraphics[width=\linewidth]{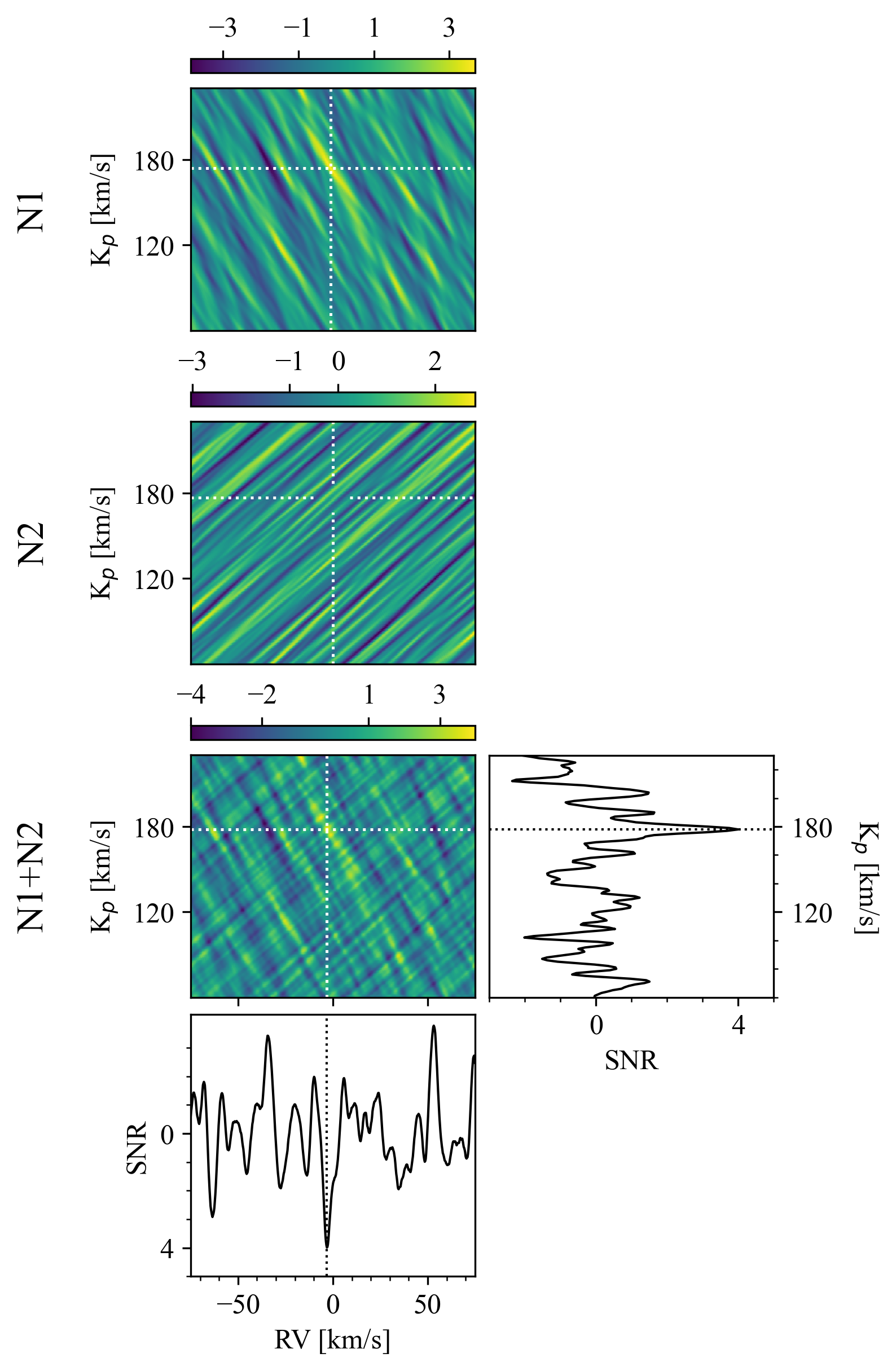}
   \caption{Top: The results of cross correlating our night 1 data with the Si model. The blue arm of the spectrograph data has been excluded from this analysis, due to the fact that it does not contain strong Si lines (see Section \ref{sec:results}). {The location of the peak $K_p$ and RV values are indicated by converging dotted white lines in each plot with a significant detection. For plots without a significant detection, the expected planetary signal location is indicated by non-converging dotted white lines (i.e., RV $= 0$ km/s, $K_p \sim 176.5$ km/s).} The colorbars indicate the same as described in the caption of Figure \ref{fig:Fe}, {though note the difference in scales}. Middle: The same as the top, but for night 2. Bottom: The same as the top, but for nights 1 and 2 combined, with each night weighted by its respective average SNR. Surrounding this plot are slices of the 2D $K_p-$RV maps at the peak $K_p$ and peak RV for the combined results. The dashed black line represents the combined peak RV and $K_p$ in each plot.}
    \label{fig:si}
\end{figure}

Of the non-detections, we note that the cross-correlation of our data with the Si model yields a peak correlation of 3.7$\sigma$ for night 1 in the red arm only (see Figure \ref{fig:si}). Although the highest correlation peak for night 2 is not located at/near the expected position, {in the final combination of both nights data}, we obtain a peak correlation of 4.0$\sigma$ at a Keplerian orbital velocity of $K_p = 178.0^{+2.2}_{-2.1}$ km/s and a radial velocity of RV$ = -3.1^{+1.8}_{-1.3}$ km/s, where we have taken as error the 1$\sigma$ extent of the correlation peak. The results of this cross-correlation for Si are presented in Figure \ref{fig:si}. Due to the fact that this peak correlation occurs near the expected location of the planetary signal, as well as the possible detection of SiO reported by \cite{Lothringer22}, it may be worth further investigating the presence of Si and its oxides in the planet's atmosphere in future works making use of additional observations. We discuss this in more detail in Section \ref{sec:discussion}.

\begin{figure*}
    \centering
    \includegraphics[width=\linewidth]{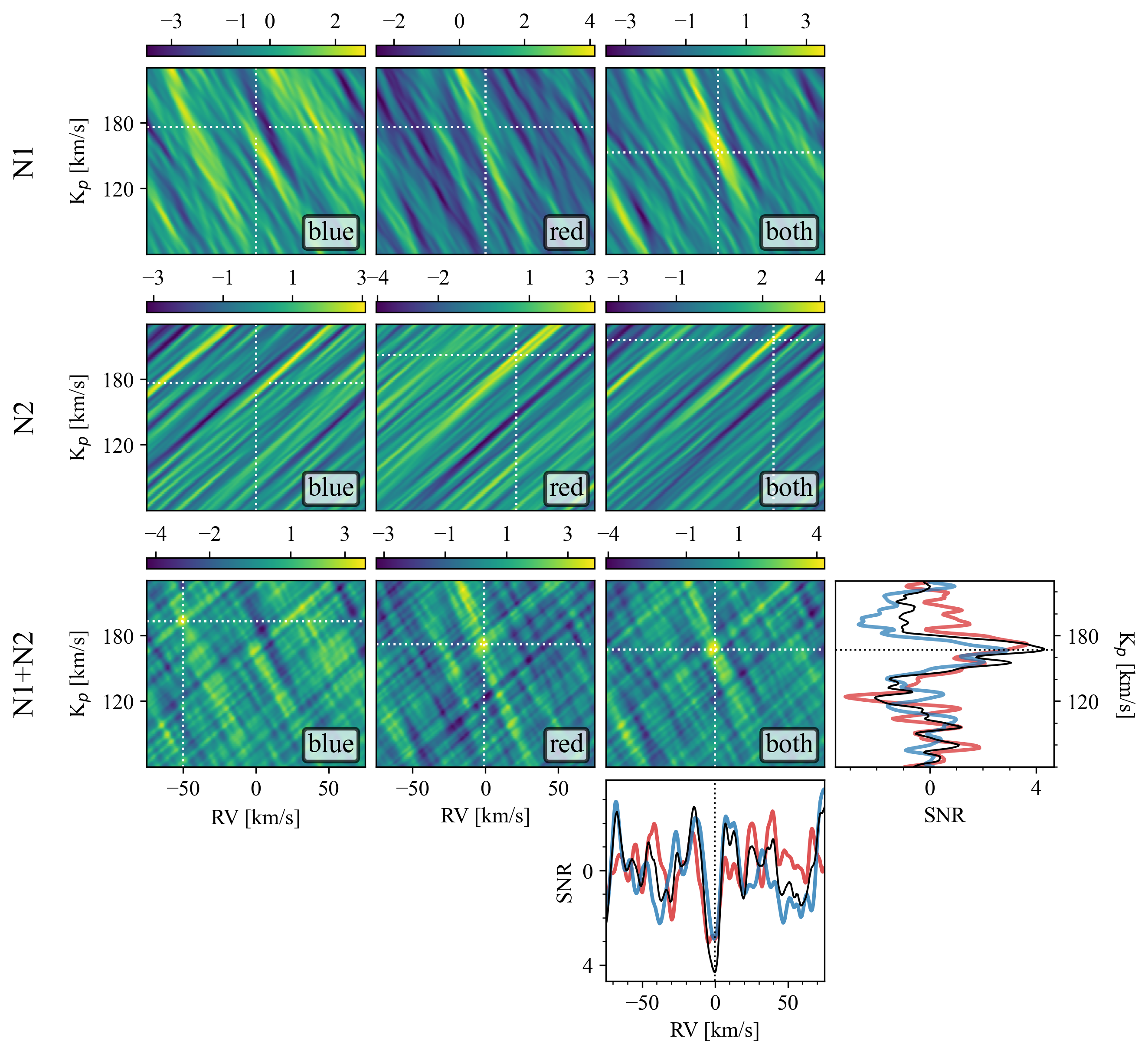}
    \caption{The results of cross-correlating our data with the Ca model. The subplots are as described in the caption of Figure \ref{fig:Fe}, though note that the colorbar scales are not the same as in Figure \ref{fig:Fe}. Although the individual nights/arms do not yield a peak correlation near the expected location of the planetary signal, we find that the combination of the blue and red arms in the first night of observations (top-right plot) along with the combination of both cameras/both nights (bottom-right plots) yield tentative detections $>3\sigma$ near the expected planetary location. In particular, the final combined significance of the correlation peak is 4.3$\sigma$ at  $K_p = 167.0^{+6.0}_{-2.9}$ km/s and RV = 0.1${}^{+2.0}_{-4.8}$ km/s.
}
    \label{fig:ca_tentative}
\end{figure*}

Likewise, we note a peak near the expected planetary location in the combined $K_p-$RV plots for Ca, as seen in Figure \ref{fig:ca_tentative}. While the signal is not seen in any individual CCF map from either night, it is visible in the combination of the red and blue cameras on night 1, and in the final combination of both cameras across both nights. The signal is present at a significance of 4.3$\sigma$ and a peak location of $K_p = 167.0^{+6.4}_{-3.1}$ km/s and RV = $0.1^{+2.3}_{-4.5}$ km/s (though note that this $K_p$ is lower than the expected value, and lower than that of our Fe detection and tentative Si detection). It is perhaps surprising that we tentatively detect Ca but do not detect Ca${}^+$, as the dayside atmosphere of WASP-178b should be hot enough to at least partially ionize any Ca present in the atmosphere. We discuss this point in additional detail in Section \ref{sec:discussion}.

\subsection{Results of the Model Injection/Recovery Test}

In general, the model injection/recovery tests yielded non-detections of injected models for the majority of species investigated in this work. This likely indicates that the atmospheric species in question are too weak to be detected given the noise level of the data, assuming that the models are indeed an accurate representation of the species in question. This is in line with the fact that, as presented in Section \ref{subsec:nondetections}, we were unable to detect the majority of species we searched for in this analysis.

\begin{figure*}
    \centering
    \includegraphics[]{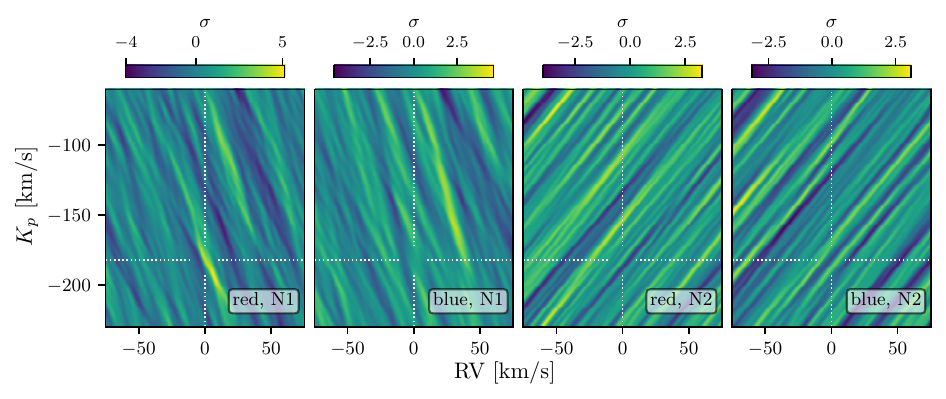}
    \includegraphics[]{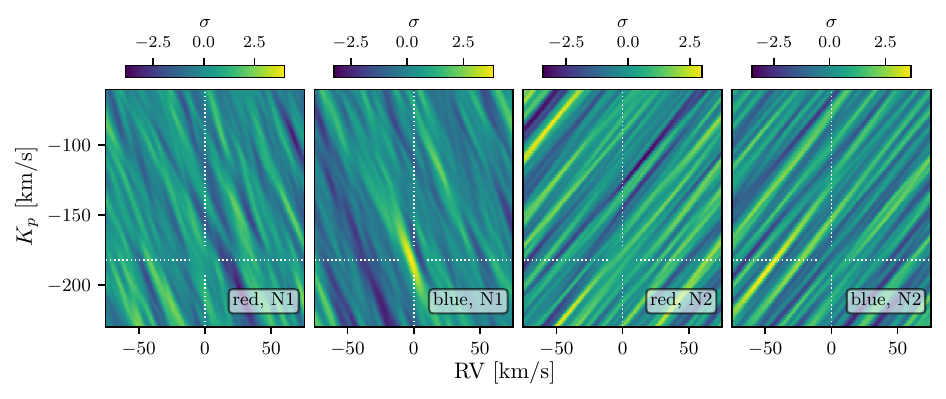}
    \caption{Top row: The results of the model injection/recovery test, as described in Section \ref{subsec:injection}, for the atmospheric model generated for Fe. The four subplots display the results for the red and blue arms of the night 1 (first two panels) and night 2 (last 2 panels) data respectively. The dashed white lines indicate the location at which the model was injected, i.e., where we expect to recover the injected signal, and the colorbars indicate the strength of the detected signal. The model was injected at a negative velocity in order to avoid boosting any real signal that was present in the data. In the case of Fe, we are only able to recover the injected model in the red arm of the night 1 data, at a significance of 5.1$\sigma$.
    Bottom row: The same as the top row, but for Ca. In this case, we are only able to recover the injected model in the blue arm of the night 1 data at 3.8$\sigma$.}
    \label{fig:fe_and_ca_injections}
\end{figure*}

The exceptions to this were the model injection/recovery tests for Fe and Ca. The results for both species are displayed in Figure \ref{fig:fe_and_ca_injections}. In the case of Fe, we are only able to recover the model that was injected into the red arm of the night 1 data. The model is recovered at a significance of 5.1$\sigma$, which is weaker than our actual detection strength in the red arm of the night 1 data (7.7$\sigma$). Given that the injected model should, in theory, be easier to detect (as we are essentially correlating the model with itself), this indicates that our 1D atmospheric models are likely not a good match for the true atmospheric signal in the data. This is perhaps not surprising, given the fact that NLTE effects may have a demonstrable effect on the atmosphere \citep{Fossati25} and we are not including these in our models. We discuss this in greater detail in Section \ref{sec:discussion}, and note that while a comparison with models including NLTE effects would be insightful, this is beyond the scope of the present work.

In the case of Ca, we find that we are tentatively able to recover the injected signal at a significance of 3.8$\sigma$ in the blue arm of the night 1 data only. Although the tentatively detected signal is not recovered at a significance of $>5\sigma$, the signal appears at the expected location and approaches $5\sigma$, therefore increasing our confidence that we are indeed recovering the injected signal. This is in line with the fact that we do tentatively recover a Ca signal in our cross-correlations (see Figure \ref{fig:ca_tentative}), albeit not in the blue arm alone. We note that, while Ca might be expected to at least partially ionize to Ca${}^+$ in the atmosphere (see Figure \ref{fig:ca_model}), our model injection/recovery tests indicate that we do not expect to detect Ca${}^+$ given the quality and amount of our data. We discuss this in additional detail in Section \ref{sec:discussion}. We also note that the strong Ca${}^+$ H \& K lines are expected to be broadened, and are located in a region of our spectra where the data quality is lower (i.e., near the edge of the ``usable'' GHOST wavelength range), which may further explain why we are unable to detect Ca${}^+$. Furthermore, while one might expect Ca to be more ionized in the hotter region of the dayside atmosphere probed by the pre-eclipse observations, our models do not take the viewing angle of the planet into account and thus cannot account for this.

\section{Discussion} \label{sec:discussion}

\subsection{Comparison with Previous High-Resolution Spectroscopy of WASP-178b}

As discussed in Section \ref{sec:intro}, WASP-178b has previously been observed at high spectral resolution via optical transmission spectroscopy \citep{Damasceno2024} and near-infrared (NIR) dayside emission spectroscopy \citep{cont2024}. As in \cite{Damasceno2024}, we are able to detect neutral iron in the planet's atmosphere, suggesting that at least some fraction of the iron present in the terminator region of the atmosphere persists in the hotter dayside region probed by the present analysis (despite the fact that the iron may at least partially ionize at these hot temperatures). Indeed, \cite{Damasceno2024} also detected ionized iron on the terminator region of the planet; while we do not detect ionized iron in the dayside, this is unsurprising based on the fact that our model injection/recovery test indicated that our data are insufficient for recovering an ionized iron model. Likewise, we are unable to detect Mg and Na (which were detected by \citealt{Damasceno2024}) for similar reasons, though it is also possible that these species may be ionized on the hotter dayside.

Our iron detection is consistent in $K_p$ with that of \cite{Damasceno2024}, though we note that the signal shown in \cite{Damasceno2024} is particularly broadened and thus encompasses a wide range of $K_p$ phase space (\citealt{Damasceno2024} discuss the possibility of dynamics contributing to their broadened signal). In terms of RV, we note a slight blueshift, though our signal is consistent within error with an offset of 0 km/s. \cite{Damasceno2024} report a blueshift in their Fe detection, consistent with possible day-night winds at the terminator region.

Our results, as well as those of \cite{Damasceno2024}, are in contrast to the analysis presented by \cite{Lothringer22}, who did not find evidence for Fe absorption in the NUV spectrum. It may be the case, as discussed in \cite{Damasceno2024}, that the region probed by the NUV observations is dominated by absorption due to Mg, Fe${}^+$, and/or SiO. Additional NUV observations may help to identify Fe in the data.

Also interesting is the fact that \cite{cont2024} do not detect neutral iron in the dayside atmosphere, despite making strong detections of ${}^{12}$CO and H${}_2$O. However, they explain this by the fact that their K-band observations do not contain a significant number of strong Fe lines. On the other hand, Fe lines are much more prominent in the optical-NIR wavelength range covered by GHOST. Indeed, \cite{cont2024} are unable to detect an injected Fe signal, meaning that Fe is not ruled out by their analysis.

Our Fe detection appears to be consistent in terms of $K_p$ with the molecular detections presented in \cite{cont2024}, though we note that their signals are relatively broad, due partially to the fact that they are only targeting post-eclipse orbital phases, and the signals from ${}^{12}$CO and H${}_2$O are themselves offset from each other with only marginal overlap. The fact that we observe Fe in emission is also consistent with the presence of a thermal inversion in WASP-178b's atmosphere, as first demonstrated by \cite{cont2024}. 

\subsection{Tentative Detections}

The hint of Si emission in the atmosphere suggested by our analysis is in line with the possible detection of SiO in the terminator region presented by \cite{Lothringer22}. If SiO is indeed responsible for the large NUV absorption depth, it may be the case that the molecule dissociates on the hotter dayside atmosphere, leading to the presence of neutral Si. Si emission has been previously detected in the dayside atmospheres of several other UHJs comparable in temperature to WASP-178b: namely, WASP-33b and KELT-20b \citep{Cont22}, as well as KELT-9b \citep[][though the Si emission appears to be under-abundant compared to the model expectations of the latter]{RH23,Zhang26}, and WASP-189b \citep{Lennart25, Sanchez25}. WASP-178b falls between the hottest and coolest of these planets in terms of equilibrium temperature, placing it in a unique region of the temperature parameter space that may allow for dissociation of SiO to be probed between the terminator and the dayside atmosphere, and where neutral Si can be present in gaseous form in the hottest regions. The presence of atmospheric Si in UHJ daysides has also been suggested by theoretical work, {with gaseous silicon-bearing species being likely precursors to silicate-based cloud particles in hot exoplanet atmospheres} \citep[e.g.,][]{Helling2019}. Our tentative detection appears to be driven by the night 1 observations, which could also indicate that we are tracing the hotter dayside region probed by the night 1 pre-eclipse observations compared to the night 2 post-eclipse data, though it is difficult to say for certain as night 2 also had poorer data quality. If indeed the signal is stronger in the pre-eclipse orbital phases than the further post-eclipse orbital phases covered in this work, it could be the case the Si is mainly localized to the planet's hotspot, which is expected to be shifted eastward, i.e., toward the evening region probed by our night 1 observations \citep[e.g.,][]{Fortney2021}. Our night 2 post-eclipse observations are closer to quadrature, where the hotspot is likely not as prominently viewed. We note as well, however, that the mostly optical wavelength coverage of GHOST is not ideal for searching for Si, which has stronger lines present in the NIR. Future work investigating the presence of Si with NIR observations of the dayside could provide additional insights into its presence in WASP-178b's atmosphere.

\begin{figure*}
    \centering
    \includegraphics[width=\linewidth]{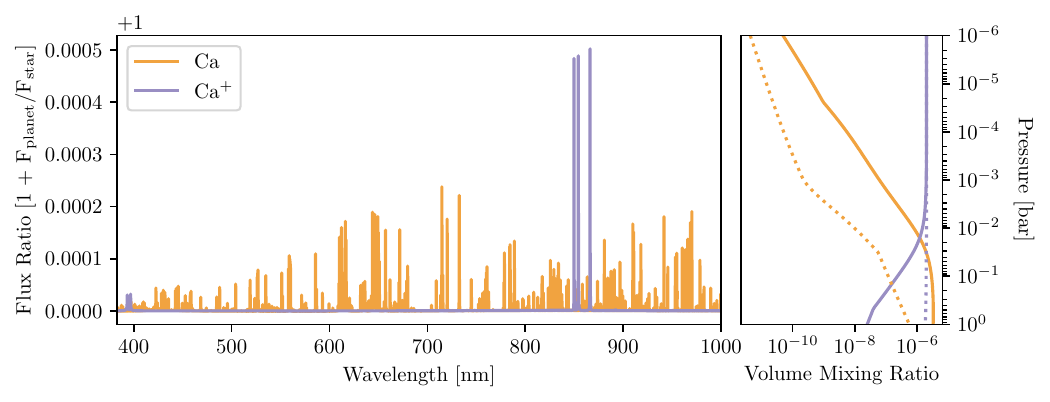}
    \caption{Left: The Ca and Ca${}^+$ models used in our analysis, as described in Section \ref{sec:models}. Right: The volume mixing ratios as a function of pressure for Ca and Ca${}^+$, as calculated by \texttt{FastChem}. The solid lines represent the VMRs calculated for WASP-178b, while the dashed lines represent possible VMRs for a hotter, KELT-9b-like planet, for comparative purposes. The two-point T-P profile for the KELT-9b-like planet is set at ($T_1$, $p_1$) = (5000 K, $10^{-3}$ bar) and ($T_2$, $p_2$) = (4000 K, $10^{-1.5}$ bar). This is estimated from the retrievals presented in \citealt{Zhang26} (their Figure 5), but is not meant to be a fully accurate representation of the atmosphere of KELT-9b. We see that at a pressure slightly below $10^{-2}$ bar, Ca begins to become more abundant than Ca${}^+$ for WASP-178b. Our observations may be probing at or around this ionization boundary. In comparison, this boundary occurs much deeper in the atmosphere for the KELT-9b-like planet.}
    \label{fig:ca_model}
\end{figure*}

Alongside our tentative detection of Si, we also report tentative hints of neutral Ca emission, as seen in Figure \ref{fig:ca_tentative}. Interestingly, however, we do not report a corresponding detection of Ca${}^+$, despite the fact that we might expect Ca to be at least partially ionized on the hot dayside of WASP-178b. We note, however, that we did not recover a detection of Ca${}^+$ in our model injection/recovery tests either. In other words, we do not expect our data to be of sufficient quality to yield a detection of Ca${}^+$ (at least at the abundance suggested by our models). This is likely due to the fact that there are few strong Ca${}^+$ lines present in the usable GHOST wavelength range, aside from the Ca${}^+$ triplet at $\sim$850 nm (as can be seen in Figure \ref{fig:ca_model}). Our analysis therefore does not rule out the presence of Ca${}^+$, but instead suggests that additional observations are needed to detect it in the atmosphere of WASP-178b. Transmission spectroscopic observations may also be useful in investigating the presence of Ca${}^+$, as they trace a higher altitude in the atmosphere where these lines may originate \citep[e.g.,][see also Figure \ref{fig:ca_model}]{Langeveld25,Zhang26}.

Both Ca and Ca${}^+$ have been previously detected in the dayside atmospheres of several UHJs. In particular, Ca was detected in the atmosphere of KELT-20b via emission spectroscopic observations from the Potsdam Échelle Polarimetric and Spectrographic Instrument (PEPSI; \citealt{Bonidie24}); and in the atmosphere of WASP-121b via emission spectroscopy from ESPRESSO \citep{Hoeijmakers24}. These were both accompanied by non-detections of Ca${}^+$; in the case of the former, this is explained by the fact that Ca${}^+$ does not have strong spectral lines in the PEPSI wavelength range, while in the case of the latter we note that while ESPRESSO covers the Ca${}^+$ H \& K lines, it does not cover the Ca${}^+$ triplet. \cite{Guo26} presented a tentative detection of Ca alongside a detection of Ca${}^+$ in the dayside atmosphere of HAT-P-70b, while \cite{RH23} and \cite{Zhang26} detected Ca${}^+$ in the dayside atmosphere of KELT-9b (alongside a non-detection of Ca in the case the former, and a detection of Ca in the latter). We note that Ca${}^+$ has also been detected via transmission spectroscopy in all of these planets' atmospheres \citep[e.g.,][]{CB19,Yan19,Nugroho20,BA20,Turner20,Borsa21,Maguire23,Darpa24,Prinoth25,Langeveld25}

In the case of our observations, we suspect that if our tentative Ca signal is real, Ca${}^+$ is likely present in the atmosphere as well but simply not detectable given the SNR of our observations. It may also be the case that Ca${}^+$ exists higher up in the atmosphere (as suggested by various transmission spectroscopic studies of Ca${}^+$; e.g., \citealt{Langeveld25}) than we are probing with our emission spectroscopic observations. This can be seen in Figure \ref{fig:ca_model}, where we plot the model atmospheres generated for Ca and Ca${}^+$ alongside the VMRs for each species calculated via \texttt{FastChem} (as described in Section \ref{sec:models}), both for WASP-178b as well as a much hotter, KELT-9b-like planet. Our emission observations are likely probing pressures around the ionization boundary where Ca starts to become more abundant than Ca${}^+$. This is also likely the case for the cooler UHJs for which Ca but not Ca${}^+$ has been detected in emission (KELT-20b, \citealt{Bonidie24}, and WASP-121b, \citealt{Hoeijmakers24}), though additional observations are needed to better constrain the presence of Ca${}^+$. Likewise, for the hotter UHJs HAT-P-70b and KELT-9b for which Ca${}^+$ has been detected on the dayside, the VMRs calculated by \texttt{FastChem} suggest that the ionization boundary between Ca and Ca${}^+$ occurs much deeper in the atmosphere, leaving Ca${}^+$ as the more abundant---and perhaps more detectable---species at the pressures probed by high-resolution emission spectroscopy.

We note as well that our tentative Ca signal is recovered at a slightly lower $K_p$ than our Fe detection, with $K_p = 167.0^{+5.9}_{-2.9}$ km/s. This may indicate that we are probing a different layer of the atmosphere tracing a different dynamical regime, but as the signal is only tentative, we cannot say for certain. Additional observations would help resolve the question of whether Ca and Ca${}^+$ are indeed present and detectable in the dayside atmosphere.

\subsection{NLTE Effects}
Our detection of iron in emission in the dayside atmosphere of WASP-178b is in line with many previous studies of ultra-hot Jupiter atmospheres \citep[e.g.,][among many others]{RH23,Deibert24,Sanchez25,Lennart25,Zhang26}. It is interesting, however, that the strength of the detected signal appears to be higher than that suggested by our 1D atmospheric models, as indicated by our model injection/recovery tests. Our models are likely under-estimating the abundance of iron present in the atmosphere, or the temperature of the atmosphere probed by our observations, or both.

Recently, \cite{Fossati25} investigated the nature of the T-P profile for WASP-178b. They found that the T-P profile is strongly driven by NLTE effects. While their analysis focused on the NUV, and on transmission spectroscopy in particular, this could perhaps also explain our lack of detection in our model injection/recovery tests, as our models are generated in an LTE framework. For example, \cite{Fossati25} find that, especially at lower pressures, NLTE effects may result in significantly higher temperatures in the atmosphere than expected from LTE models (see their Figure 2). They also found that their LTE transmission spectra tended to understimate absorption, due to the models being cooler than their NLTE transmission spectra. While we are focusing on emission in this work, the same may be true in our case. Although we are able to detect Fe regardless of the fact that NLTE effects are not included in our models, future work could focus on investigating the presence of atmospheric species with NLTE models similar to those presented in \cite{Fossati25}, but for emission spectra specifically.

\subsection{Future Directions}

{Although a full atmospheric retrieval \citep[e.g.,][]{BrogiLine2019} is beyond the scope of the present work, future analyses may benefit from investigating these observations in a retrieval framework.}

{Future analyses may also benefit from a joint optical-NIR analysis including both the GHOST observations presented in this work as well as the CRIRES+ NIR dayside observations presented in \cite{cont2024}. While \cite{cont2024} were able to constrain the presence of various volatile species in the planet's atmosphere (e.g., ${}^{12}$CO and H${}_2$O), our GHOST observations provide additional constraints on the planet's atmospheric refractory content, including detections of refractory species which could not be recovered at the CRIRES+ wavelength range used in \cite{cont2024}. Analyzed together, these two datasets may therefore allow for measurements of the planet's refractory-to-volatile ratio, which could shed light on its formation pathway \citep[e.g.,][]{Lothringer21, Chachan23}, following analyses such as \cite[e.g.,][]{Smith24, Sanchez25, Kanumalla26},}

\section{Conclusions}
\label{sec:conclusions}
In this work, we presented high-resolution optical dayside emission spectroscopy of the ultra-hot Jupiter WASP-178b obtained with GHOST at the Gemini South Observatory. We observed the planet over the course of two 4.5 h observing periods, covering pre- and post-eclipse orbital phases. We then searched for a suite of molecular and atomic species via the Doppler cross-correlation technique, making a strong detection of Fe emission in the atmosphere alongside tentative detections of neutral Si and Ca emission.

If the neutral Si signal suggested by our data is in fact real, it adds a new data point to previous detections of Si in the dayside atmospheres of other UHJs \citep[e.g.,][]{Cont22,RH23,Zhang26}, highlighting the importance of silicon chemistry in these planets' atmospheres. {Given the likely importance of silicates in forming cloud particles in hot giant exoplanet atmospheres \citep[e.g.,][]{Helling2019}, the detection of gaseous Si on the planet's dayside may help trace the high-temperature regime under which silicon-bearing species remain gaseous, rather than condensing into clouds or cloud precursors.} Finally, this tentative detection also adds further context to the previous inference of SiO in the planet's terminator region made by \cite{Lothringer22} with HST/WFC3/UVIS data, suggesting possible dissociation of the molecule in the hottest regions of the dayside atmosphere. Future observations, including in particular those at NIR wavelengths, will help shed light on this tentative detection, {though our work demonstrates the potential of red-optical spectroscopy in further constraining silicon chemistry and the gas-phase precursors to silicate cloud formation in ultra-hot exoplanet atmospheres. Observations of the planet's terminator region via high-resolution transmission spectroscopy may also reveal whether atmospheric Si persists in the morning/evening region of the planet, or whether it may have begun to be sequestered into Si-bearing molecules and/or condensates as temperatures decrease towards the planet's nightside.}

Our work adds to the growing body of literature on the ultra-hot Jupiter WASP-178b, which orbits one of the hottest stars known to host a UHJ, shedding light on the planet's 3D atmospheric chemical processes. Our high-resolution optical dayside observations also build upon previous optical studies of WASP-178b's terminator region as well as NIR observations of its dayside, helping to map its 3D physical and chemical processes. Future work will benefit from more sophisticated modelling techniques (possibly incorporating NLTE effects), as well as additional observations at a range of orbital phases and wavelength regimes in order to better understand the extreme atmosphere of this ultra-hot world. 

\begin{acknowledgements}
{We thank the referee for their careful and constructive review of this manuscript, which has helped strengthen this work.}

    Based on observations obtained at the International Gemini Observatory, a program of NSF NOIRLab, which is managed by the Association of Universities for Research in Astronomy (AURA) under a cooperative agreement with the U.S. National Science Foundation on behalf of the Gemini Observatory partnership: the U.S. National Science Foundation (United States), National Research Council (Canada), Agencia Nacional de Investigaci\'{o}n y Desarrollo (Chile), Ministerio de Ciencia, Tecnolog\'{i}a e Innovaci\'{o}n (Argentina), Minist\'{e}rio da Ci\^{e}ncia, Tecnologia, Inova\c{c}\~{o}es e Comunica\c{c}\~{o}es (Brazil), and Korea Astronomy and Space Science Institute (Republic of Korea). Observation ID: GS-2024A-Q-138; PI: Deibert.

    We thank Joy Chavez, Daniel May, German Gimeno, and Pablo Candia for their work in obtaining the observations presented in this paper.

    This work is supported by the U.S. National Science Foundation through their Research Experiences for Undergraduates (REU) program under Award No.\ 2349023. We thank Joan Najita, Guillermo Damke, Aleksandar Cikota, Pol Massana, Nicole Auza, and the staff of the Gemini South Observatory and NSF/NOIRLab in La Serena for their work in organizing the REU program. We also thank John Blakeslee for the support.

    EKD acknowledges the support of a Banting Postdoctoral Fellowship - NSERC.

    E.dM. would like to acknowledge support from the UK Science and Technology Facilities Council (STFC, grant number ST/X00094X/1).

    J.D.T. acknowledges funding support for this research by the TESS Guest Investigator Program G06165.

\end{acknowledgements}

\facilities{Gemini:South (GHOST), Exoplanet Archive}

\software{\texttt{Astropy} \citep{astropy:2013, astropy:2018, astropy:2022}, \texttt{DRAGONS} \citep{dragons19,dragons22} \texttt{FastChem} \citep{Stock18}, \texttt{Matplotlib} \citep{Hunter:2007}, \texttt{NumPy} \citep{numpy}, \texttt{petitRADTRANS} \citep{petitradtrans}, \texttt{SciPy} \citep{scipy}}

\appendix

\section{\textsc{SysRem} Results}
\label{app:sysrem}
\begin{sidewaysfigure}
    \begin{flushleft}
    \includegraphics[width=\textheight]{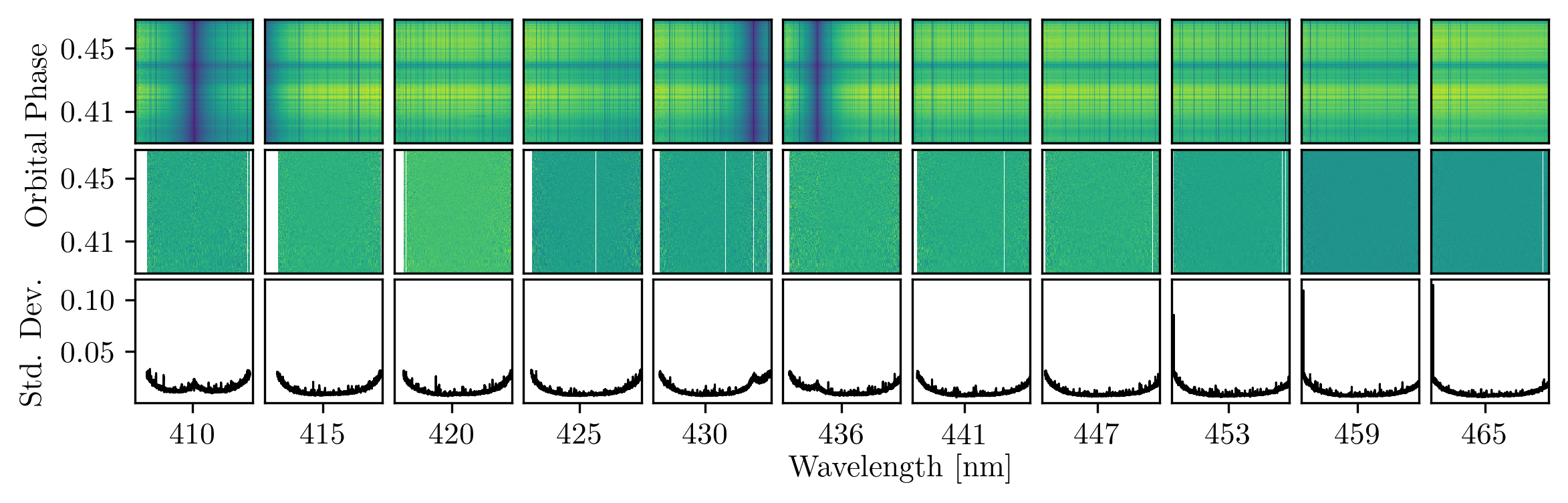}
    \includegraphics[width=0.917\textheight]{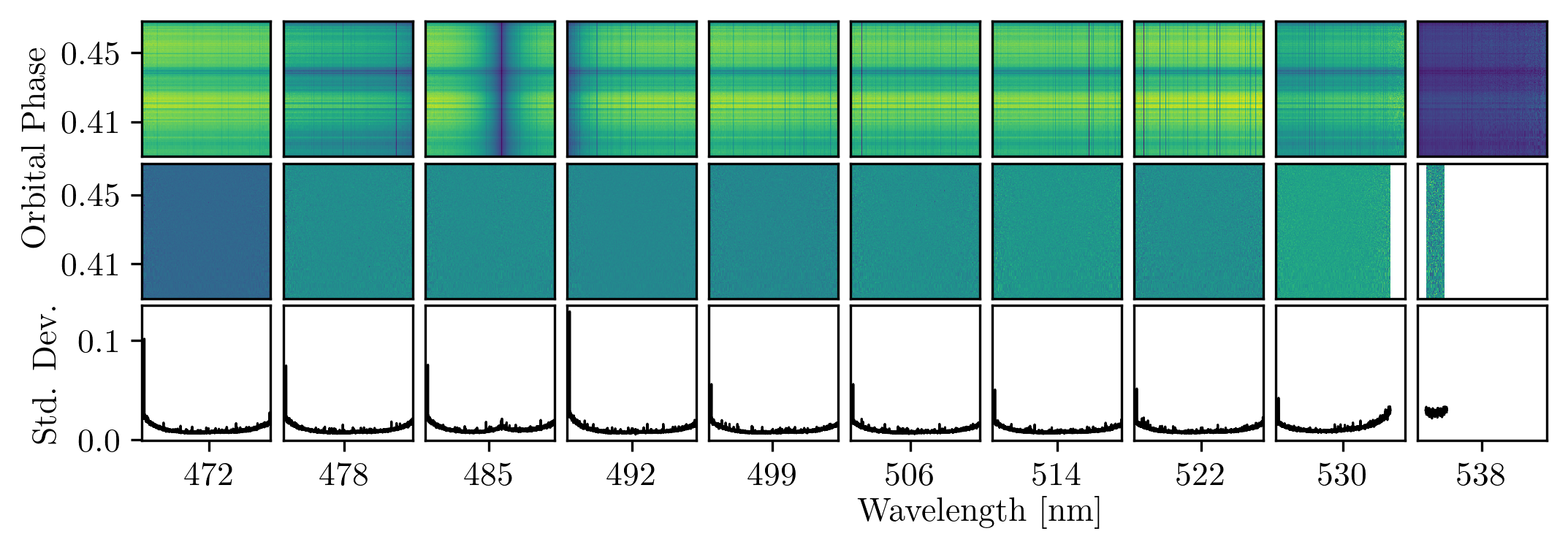}
    \end{flushleft}
    \centering
    \caption{The results of applying the SysRem algorithm to the first 11 orders (top three rows) and last 10 orders (bottom three rows) in the night 1 blue arm of the spectrograph. Top row: The data reduced by DRAGONS. Note that there may be some overlap in wavelength between subsequent orders; each order is displayed separately. Middle row: The results of applying 6 iterations of SysRem to the data as described in Section \ref{subsec:sysrem}. At this point, the telluric and stellar absorption features have been removed, and the planetary signal is buried in the noise. Bottom row: The standard deviation of the data after applying SysRem. The y-scale is the same for all plots in this row.}
    \label{fig:20240324_blue_sideways_1&2}
\end{sidewaysfigure}

\begin{sidewaysfigure}
    \begin{flushleft}
    \includegraphics[width=\textheight]{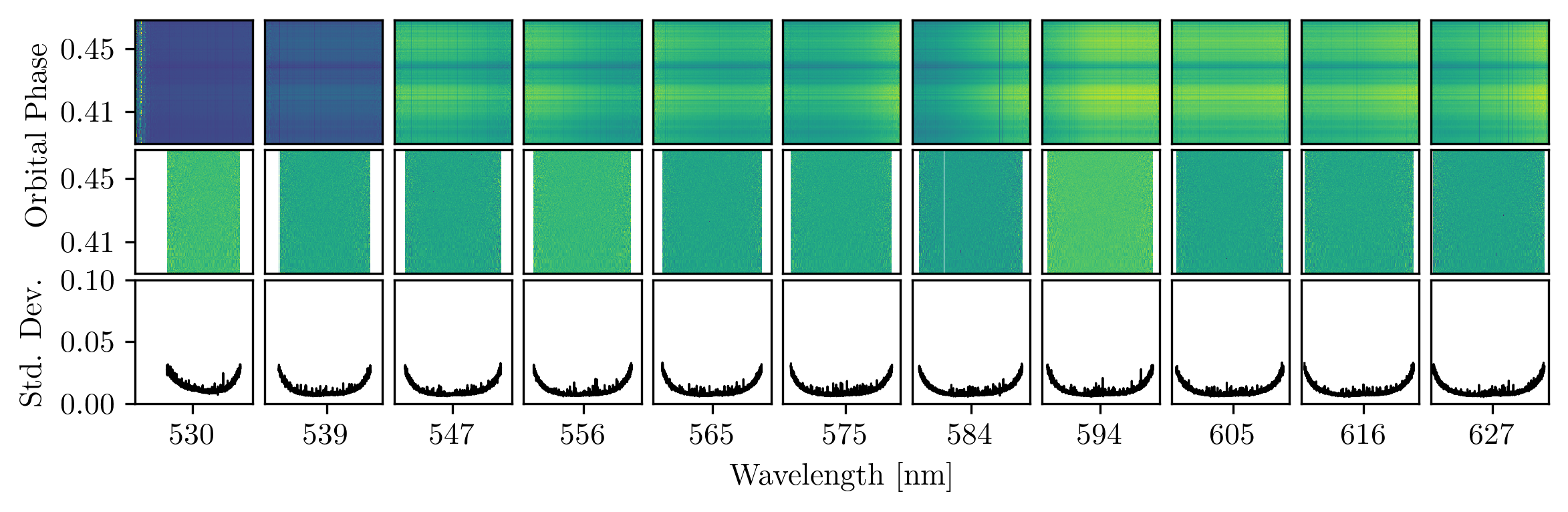}
    \includegraphics[width=\textheight]{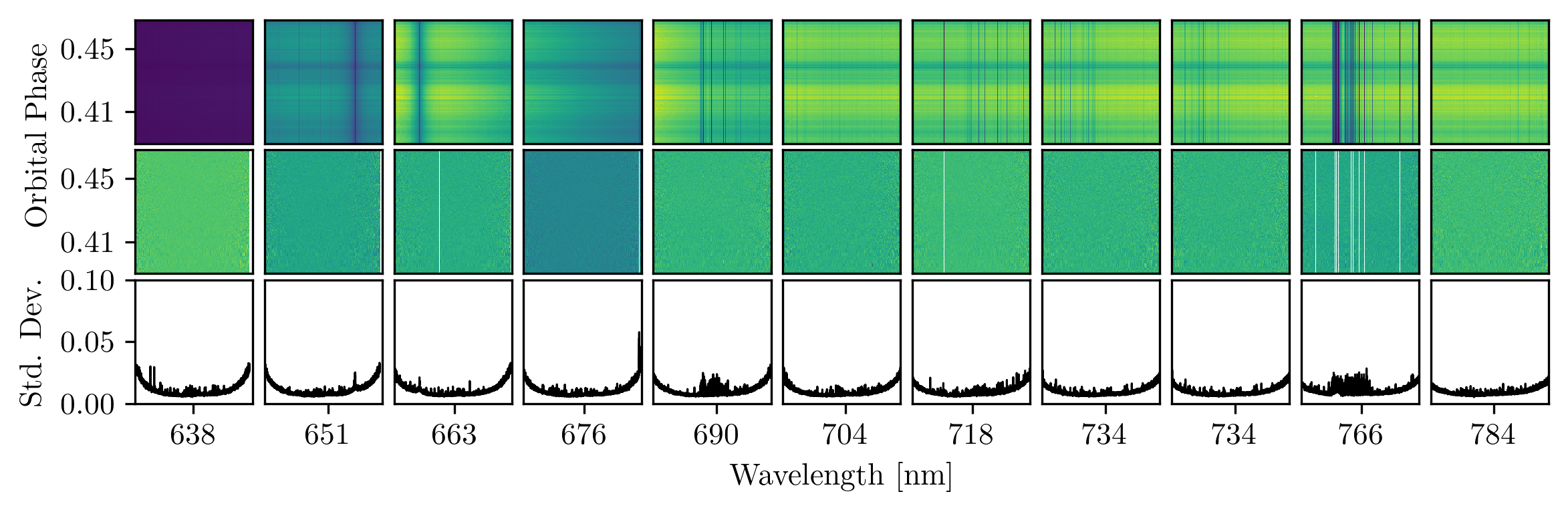}
    \end{flushleft}
    \centering
    \caption{The same as Fig. \ref{fig:20240324_blue_sideways_1&2}, but for the first 11 orders (top three rows) and next 11 orders (bottom three rows) in the night 1 red arm of the spectrograph. As explained in Section \ref{subsec:sysrem}, we applied 3 iterations of SysRem to the red data}
    \label{fig:20240324_red_sideways_1&2}
\end{sidewaysfigure}

\begin{figure*}
    \includegraphics{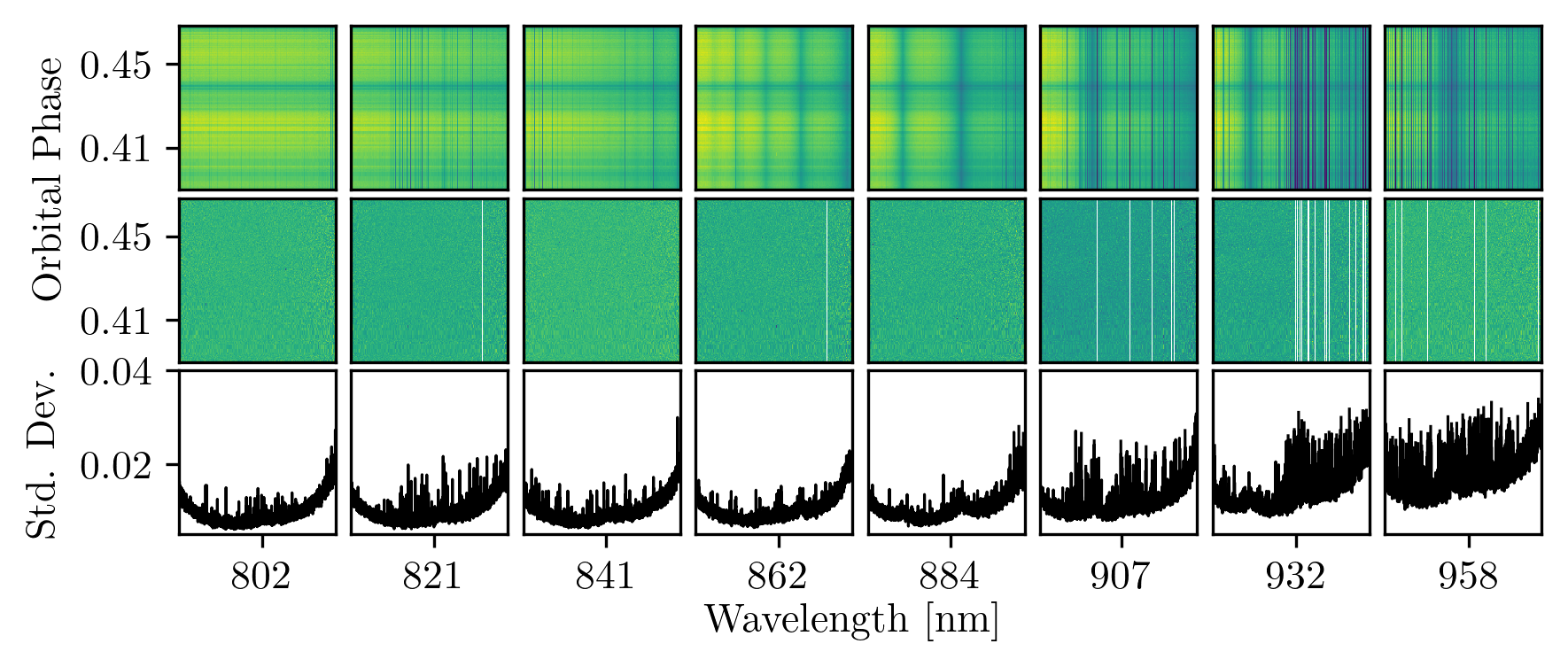}
    \centering
    \caption{The same as Fig. \ref{fig:20240324_red_sideways_1&2}, but for the last 8 orders in the night 1 red arm of the spectrograph.}
    \label{fig:20240324_red_sideways_3}
\end{figure*}

\begin{sidewaysfigure}
    \begin{flushleft}
    \includegraphics[width=\textheight]{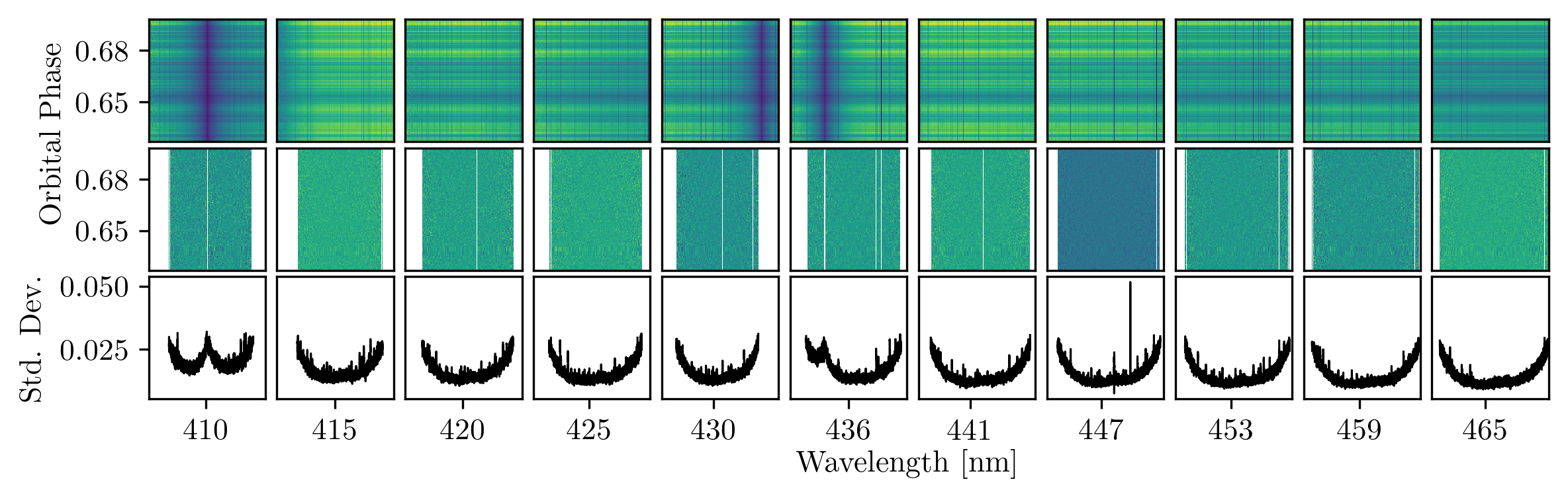}
    \includegraphics[width=0.918\textheight]{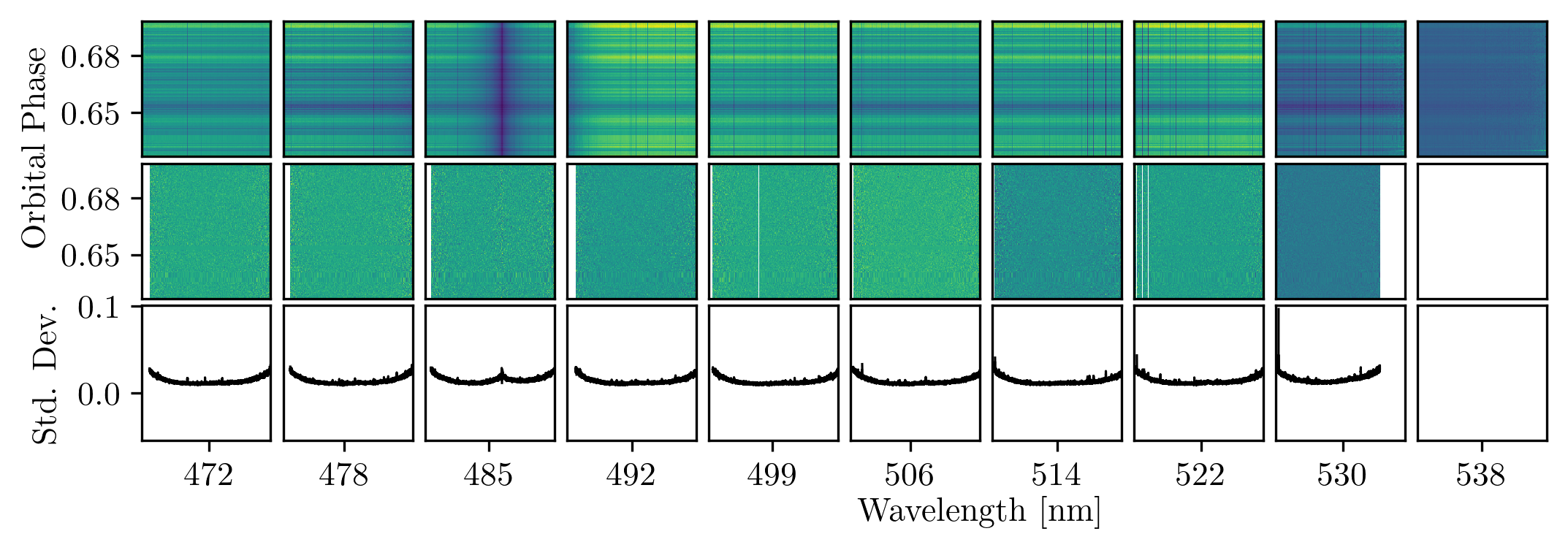}
    \end{flushleft}
    \centering
    \caption{The same as Fig. \ref{fig:20240324_blue_sideways_1&2}, but for the night 2 blue data.}
    \label{fig:20240504_blue_sideways_1&2}
\end{sidewaysfigure}

\begin{sidewaysfigure}
    \begin{flushleft}
    \includegraphics[width=\textheight]{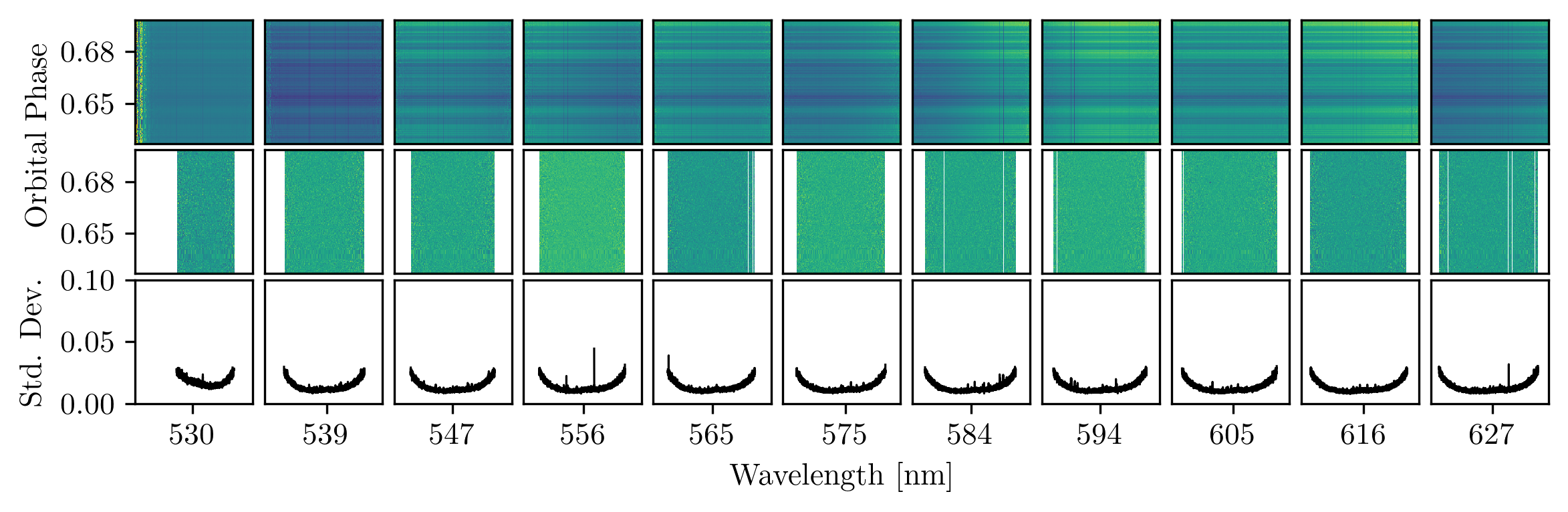}
    \includegraphics[width=\textheight]{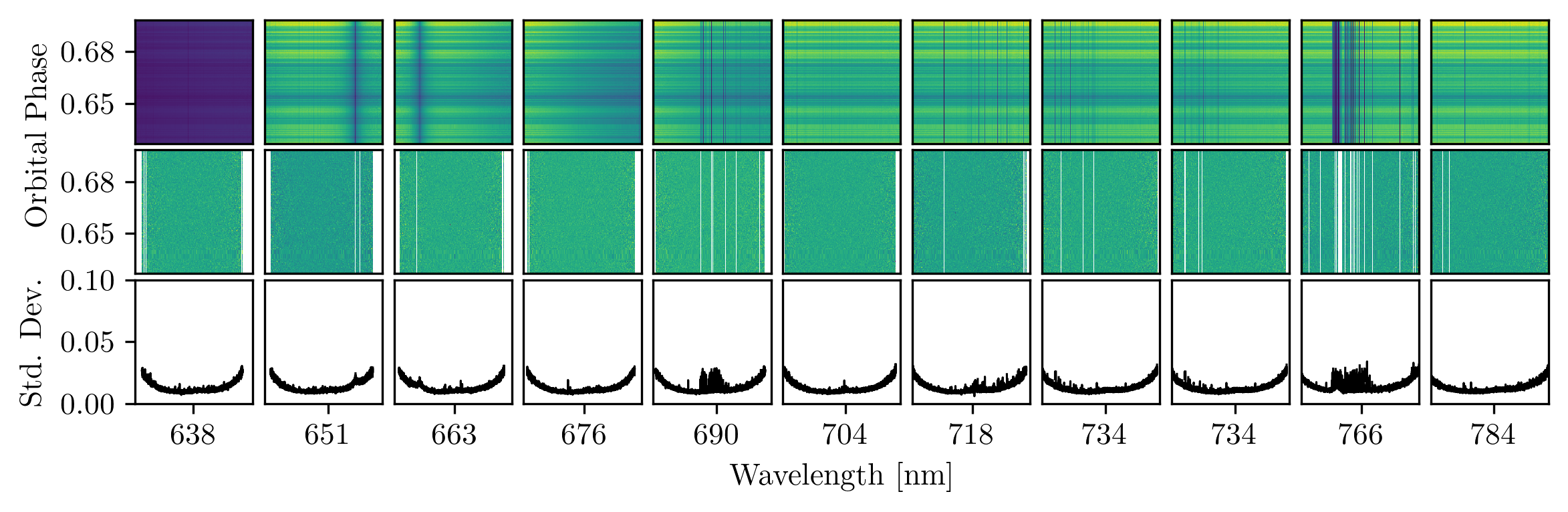}
    \end{flushleft}
    \centering
    \caption{The same as Fig. \ref{fig:20240324_red_sideways_1&2}, but for the night 2 red data.}
    \label{fig:20240504_red_sideways_1&2}
\end{sidewaysfigure}

\begin{figure*}
\includegraphics{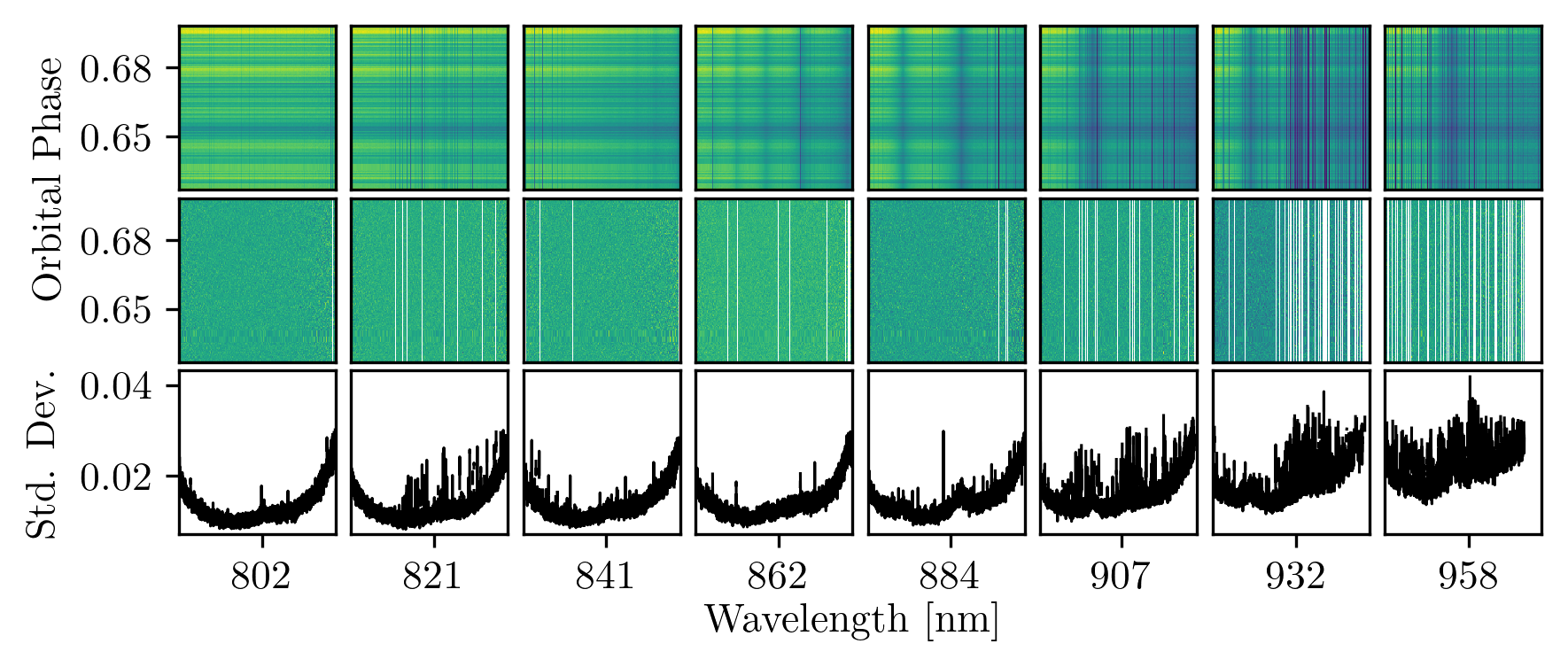}
    \centering
    \caption{The same as Fig. \ref{fig:20240324_red_sideways_3}, but for the last 8 orders in the night 2 red arm spectrograph.}
    \label{fig:20240504_red_sideways_3}
\end{figure*}

The results of applying the \textsc{SysRem} algorithm as described in Section \ref{subsec:sysrem} to our data are presented in Figs. \ref{fig:20240324_blue_sideways_1&2}, \ref{fig:20240324_red_sideways_1&2}, \ref{fig:20240324_red_sideways_3}, \ref{fig:20240504_blue_sideways_1&2}, \ref{fig:20240504_red_sideways_1&2}, and \ref{fig:20240504_red_sideways_3}. Note that this does not include the excluded orders as described in Section \ref{sec:observations}.


\bibliography{bibliography}{}
\bibliographystyle{aasjournalv7}

\end{document}